\documentclass{iopart}

\usepackage{graphicx}
\usepackage{bm}
\usepackage{sidecap}

\begin{document}

\title{One-particle dynamical correlations in the one-dimensional Bose gas}

\author{
Jean-S\'ebastien Caux $^1$,
Pasquale Calabrese $^2$
and Nikita A. Slavnov $^3$} 
\address{$^1$ Institute for Theoretical Physics, University of
  Amsterdam, 
  1018 XE Amsterdam, The Netherlands.}
\address{$^2$ Dipartimento di Fisica dell'Universit\`a di Pisa and INFN, Pisa, Italy.}
\address{$^3$ Steklov Mathematical Institute, Gubkina 8, Moscow 117966, Russia.}

\date{\today}

\begin{abstract}
The momentum- and frequency-dependent one-body correlation function of the 
one-dimensional interacting Bose gas (Lieb-Liniger model) in the repulsive regime 
is studied using the Algebraic Bethe Ansatz and numerics.  We 
first provide a determinant representation for the field form factor 
which is well-adapted to numerical evaluation.  The correlation function
is then reconstructed to high accuracy for systems with finite but
large numbers of particles, for a wide range of values of the interaction parameter. 
Our results are extensively discussed, in particular their specialization to the static case.
\end{abstract}

\maketitle

\section{Introduction}

For many decades now, there has been a sustained interest in the physics of 
one-dimensional quantum systems, in no small part due to 
Bethe's fundamental paper on the Heisenberg spin chain \cite{BetheZP71}.
This publication laid the foundation for what has today become one of the most important
fields of mathematical physics, namely the theory of Bethe Ansatz integrable models.

More recently, renewed motivation for the study of interacting quantum models in 1D has come from their growing number of
experimental realizations.  The simplest system to consider is probably the one-dimensional gas of bosons
with local (delta function) interaction (known as the Lieb-Liniger model \cite{LiebPR130}), 
which can effectively describe atoms confined using optical lattices 
\cite{GoerlitzPRL87,GreinerPRL87,Greinernat,MoritzPRL91,ParedesNATURE429,KinoshitaSCIENCE305,KoehlAPB79,FertigPRL94,kww-06}. 
A wide range of interactions are accessible, even the strongly-interacting (free fermion-like)
limit considered long ago by Tonks and Girardeau
\cite{TonksPR50,GirardeauJMP1}.

The integrability of the Lieb-Liniger model is by now textbook material
\cite{GaudinBOOK,KorepinBOOK,TakahashiBOOK,MattisBOOK},
and many of the standard methods now used in the field were in fact first developed and applied to it.
For example, after the construction of the ground state and elementary excitations in \cite{LiebPR130},
the Yang-Yang approach to thermodynamics was introduced in \cite{YangJMP10}, leading
the way to nonperturbative results on the equilibrium properties of the infinite system
at arbitrary temperatures.

The motivation of the present paper is to help further our understanding of the
interacting Bose gas in one dimension, by addressing one of the long-standing unresolved
problems associated to it:  the calculation of correlation functions.  The importance of
this is clear, in view of the fact that most experimentally-accessible response functions
are expressible in terms of a few basic dynamical correlation functions of local physical
fields.  

Because of the relative complexity of the Bethe Ansatz eigenfunctions, it has proved much
more difficult to provide information on correlation functions than on thermodynamics. 
For the interacting Bose gas, the first set of results probably dates
back to the work of Girardeau \cite{GirardeauJMP1}, who considered the
strongly-interacting limit.  This represents a major simplification in
that the spectrum becomes that of free fermions.  The bosonic density 
operator then coincides with a free fermionic one, in view of its blindness to particle
statistics.  Handling particle statistics in the same limit allows to obtain the one-body density matrix
as the determinant of a Fredholm integral operator, both at zero and nonzero temperatures
\cite{LenardJMP5,LenardJMP7,JimboPD1,KorepinCMP129}.

Away from the Tonks-Girardeau (TG) limit, static local moments of the density operator can
be obtained.  The second moment $g_2 = \langle :n^2(x): \rangle$ is given directly from
the derivative of the partition function with respect to the interaction 
parameter (the Hellmann-Feynman theorem) \cite{KheruntsyanPRL91}, but higher moments $g_n$ with $n > 2$ are more challenging.  These were
computed for arbitrary temperature in a large coupling expansion in \cite{GangardtPRL90,GangardtNJP5,bg-06}.
At zero temperature, the third moment $g_3$ was obtained exactly for any interaction
in \cite{Cheianov0506609} using an integrable lattice regularization of the Lieb-Liniger model
coupled with Conformal Field Theory \cite{BelavinNPB241,CFTBOOK} considerations.

The full calculation of correlation functions beyond either the static of local cases
remains analytically intractable to this day.  Besides formulations for the asymptotics based
on Conformal Field Theory, there exist important results 
providing determinant representations of the exact correlators in terms of auxiliary 
quantum fields, which ultimately map the
infinite-volume problem to the calculation of the determinant of a Fredholm operator
obeying a completely integrable integro-differential equation 
\cite{KorepinCMP136,KojimaCMP188,KojimaCMP189,ItsTMF119,SlavnovTMF121}.  
The asymptotics of the correlation functions can then be obtained by solving a Riemann-Hilbert
problem (see \cite{KorepinBOOK} and references therein for an introduction to the subject).
Closed-form analytical expressions for general correlation functions have however
not yet been obtained.

Here, we make use of another strategy which one
of us originally developed for dynamical spin-spin correlation functions in Heisenberg spin chains
\cite{CauxPRL95,CauxJSTATP09003}.  
Our objective is not to give final answers to the theoretical computation of 
correlation functions, but rather to provide a reliable and practical computational scheme for obtaining them
to very high precision.  This will be of great help for experimental comparisons (where finite resolutions
are inevitable) or comparison with other theoretical methods.  We specialize here
to one-body correlations, following on our earlier work \cite{CauxPRA74}
on dynamical density-density correlations.

The overall strategy is summarized as follows:  two-point correlation functions are expressed as an 
appropriate sum of form factors 
over intermediate states, and the summation is carried out numerically.  
The form factors of local operators are expressed as determinants of matrices whose
entries are simple functions of the sets
of Bethe rapidities of the two eigenstates involved (for the density, they were obtained in
\cite{SlavnovTMP79_82,KojimaCMP188};  see however the more practical representaion we offer here for 
the field form factor).  
The summation over intermediate states can itself
be aggressively optimized by making use of the fact that not all intermediate states contribute equally to
the desired correlator.
Such an optimized method is in principle applicable to any integrable model as long as the determinant representations 
of the form factors are known,
and was baptized the ABACUS method \cite{ABACUS}.
We here carry on with our study of the Lieb-Liniger model 
by providing the corresponding computations for its one-particle dynamical correlation
function.  

The plan of the paper is as follows.  We first review the basic formulation of the
theory, and define the correlations we will be interested in.  The Algebraic Bethe
Ansatz for the model is thereafter presented, and used to derive a
simplified representation for the one-particle form factor in terms of the determinant of a 
single matrix.  We then present the results for
the correlation functions for a range of values of the interaction parameter, and
discuss the results in detail, in particular their specialization to the static case
(momentum distribution function and one-body density matrix).  Conclusions and directions for future work are
presented at the end.

\section{Definitions}
We consider a ring of length $L$, with $N$ bosonic particles interacting repulsively
with each other through a local potential of strength $c$.  
The time evolution of this system is controlled by the Lieb-Liniger Hamiltonian \cite{LiebPR130}
\begin{equation}
{\cal H}_N = -\sum_{j=1}^N \frac{\partial^2}{\partial x_j^2} + 2c \sum_{1 \leq j < l \leq N} \delta(x_j - x_l).
\end{equation}
An alternate second-quantized formulation is obtained by introducing canonical 
Bose fields $\Psi (x)$ and $\Psi^{\dagger} (x)$ with equal-time commutation relations
\begin{equation}
\left[ \Psi (x), \Psi^{\dagger} (y) \right] = \delta (x-y), \hspace{0.5cm}
\left[\Psi(x), \Psi(y) \right] = \left[\Psi^{\dagger}(x) , \Psi^{\dagger}(y) \right] = 0.
\end{equation}
The Fock vacuum $| 0 \rangle$ is then defined by 
\begin{equation}
\Psi (x) | 0 \rangle = 0, \hspace{0.5cm} \langle 0 | \Psi^{\dagger} (x) = 0, \hspace{0.5cm} 
\langle 0 | 0 \rangle = 1,
\end{equation}
and thus the model corresponds to the quantum nonlinear Schr\"odinger equation,
\begin{equation}
H = \int_0^L dx \left[ \partial_x \Psi^{\dagger} (x) \partial_x \Psi (x) 
+ c \Psi^{\dagger} (x) \Psi^{\dagger} (x) \Psi (x) \Psi(x) \right].
\end{equation}
The Bethe Ansatz provides a basis for the Fock space.  More precisely,
in the repulsive case $c > 0$ which we are considering, each eigenfunction in the $N$-particle sector 
is fully described by a set of $N$ real parameters $\{ \lambda \}_N$, solution
to a set of coupled nonlinear equations (the Bethe equations, which we will 
review and discuss in more details in the next section).   

Our aim here is to present computations of the zero temperature one-particle dynamical correlation function
\begin{equation}
G_2 (x,t) \equiv \langle \Psi^{\dagger} (x, t) \Psi (0, 0) \rangle_N 
\label{G2}
\end{equation}
as a function of the interaction parameter $c$ (we consider here only the repulsive case $c > 0$).
By identifying the state $| \{ \lambda \} \rangle $ with the ground state and using 
the closure relation of Bethe eigenstates, 
this correlation function can be represented as a sum over form factors of the local field
operator between ground- and excited state as
\begin{eqnarray}
G_2 (x, t) = \sum_{\{ \mu \} } \frac{\langle \{ \lambda \} | \Psi^{\dagger} (x, t) | \{ \mu \} \rangle
\langle \{ \mu \} | \Psi (0, 0) | \{ \lambda \} \rangle}{\langle \{ \lambda \} | \{ \lambda \} \rangle \langle \{ \mu \} | \{ \mu \} \rangle}.
\end{eqnarray}
The evaluation of this expression can be performed in three steps:  first, finding the solution of the Bethe equations;
second, computing the form factors;  finally, performing the summation over intermediate
states.  We begin by providing a new representation for the form factors, using the Algebraic
Bethe Ansatz, which allows all these steps to be efficiently implemented numerically
for finite numbers of particles.  

\section{Algebraic Bethe Ansatz and Form Factors}
The central object of the Algebraic Bethe Ansatz \cite{FaddeevTMP40} (or Quantum Inverse Scattering Method; for an
introduction to the concepts and terminology, see \cite{KorepinBOOK}) is the $R$-matrix, which
solves the Yang-Baxter equation and takes the following form for the 
Lieb-Liniger model (quantum nonlinear Schr\"odinger equation):
\begin{eqnarray}
R (\lambda, \mu) = \left( \begin{array}{cccc}
f (\mu, \lambda) & 0 & 0 & 0 \\
0 & g(\mu, \lambda) & 1 & 0 \\
0 & 1 & g (\mu, \lambda) & 0 \\
0 & 0 & 0 & f(\mu, \lambda) \end{array} \right),
\end{eqnarray}
in which
\begin{equation}
g(\lambda, \mu) = \frac{ic}{\lambda - \mu}, \hspace{0.5cm} f(\lambda, \mu) = \frac{\lambda - \mu + ic}{\lambda - \mu}.
\end{equation}
Other standard functions which we will use later are defined as
\begin{equation}
h(\lambda, \mu) = \frac{\lambda - \mu + ic}{ic}, \hspace{0.5cm} 
t(\lambda, \mu) = \frac{(ic)^2}{(\lambda - \mu)(\lambda - \mu + ic)} = \frac{g(\lambda, \mu)}{h(\lambda, \mu)}.
\end{equation}
The monodromy matrix, which will be used to construct all the conserved charges (in particular,
the Hamiltonian) is represented in auxiliary space as
\begin{eqnarray}
T(\lambda) = \left( \begin{array}{cc}
A(\lambda) & B(\lambda) \\ C(\lambda) & D(\lambda) \end{array} \right),
\end{eqnarray}
where the $A,B,C,D$ operators act in the Fock space as follows.  The vacuum is an eigenvector
of $A$ and $D$,
\begin{equation}
A(\lambda) | 0 \rangle = a(\lambda) | 0 \rangle, \hspace{0.5cm} D(\lambda) | 0 \rangle = d(\lambda) | 0 \rangle
\end{equation}
with $a(\lambda) = e^{-iL\lambda/2}$ and $d(\lambda) = e^{iL\lambda/2}$, whereas $B$ and $C$ respectively
act as raising and lowering operators with the annihilation properties
\begin{equation}
\langle 0 | B(\lambda) = 0, \hspace{0.5cm} C(\lambda) | 0 \rangle = 0.
\end{equation}
The commutation relations between these operators are quadratic, and given by
\begin{equation}
R(\lambda, \mu) (T(\lambda) \otimes T(\mu)) = (T(\mu) \otimes T(\lambda)) R(\lambda, \mu).
\end{equation}
The Hamiltonian is related by trace identities to the trace of the monodromy matrix,
$A(\lambda) + D(\lambda)$, and these can therefore be diagonalized simultaneously.
The (unnormalized) eigenvectors of the Hamiltonian are constructed as
\begin{equation}
|\{ \lambda \}_N \rangle = \prod_{j=1}^N B(\lambda_j) | 0 \rangle,
\hspace{0.5cm}
\langle \{ \lambda \}_N | = \langle 0 | \prod_{j=1}^N C(\lambda_j),
\end{equation}
provided that the rapidities are solution to the Bethe equations
\begin{equation}
e^{i \lambda_j L} = \prod_{l \neq j} \frac{\lambda_j - \lambda_l + ic}{\lambda_j - \lambda_l - ic}, 
\hspace{0.5cm} j = 1, ..., N
\end{equation}
or in logarithmic form,
\begin{equation}
\lambda_j + \frac{1}{L} \sum_{l=1}^N 2 ~\mbox{arctan} \frac{\lambda_j - \lambda_l}{c} = \frac{2\pi}{L} I_j,
\hspace{0.5cm} j = 1, ..., N
\label{logBethe}
\end{equation} 
where the quantum numbers $I_j$ are half-odd integers for $N$ even, and integers for $N$ odd.
Proper eigenfunctions are obtained for sets of non-coincident rapidities $\lambda_j \neq \lambda_l, j \neq l$.
Since the left-hand side of (\ref{logBethe}) is a monotonous function of $\lambda_j$, it can
be proven that all solutions are real \cite{YangJMP10};  we can moreover span the whole Fock space by choosing
sets of ordered quantum numbers $I_j > I_l, j > l$ meaning that $\lambda_j > \lambda_l, j > l$.

Once the Bethe equations are solved for the rapidities, the eigenstate is fully
characterized.  The state norm
is obtained from the Gaudin-Korepin formula \cite{GaudinBOOK,KorepinCMP86},
which can be proven by making use of the commutation relations between $B$ and $C$ operators:
\begin{eqnarray}
\langle \{ \lambda \}_N | \{ \lambda \}_N \rangle 
= c^N \prod_{j > k = 1}^N \frac{ \lambda_{jk}^2 + c^2}{\lambda_{jk}^2}
\mbox{det}_N {\cal G} (\{ \lambda \}) \equiv || \{ \lambda \}_N ||^2
\label{norm}
\end{eqnarray}
in which ${\cal G}$ is the Gaudin matrix, whose entries are simple analytical functions of the 
rapidities,
\begin{eqnarray}
{\cal G}_{jk} (\{ \lambda \}_N) = \delta_{jk} \left[ L + \sum_{l=1}^N K(\lambda_j, \lambda_l) \right] - K (\lambda_j, \lambda_k)
\label{Gaudin}
\end{eqnarray}
where the kernel is
\begin{equation}
K (\lambda, \mu) = \frac{2c}{(\lambda - \mu)^2 + c^2}.
\label{kernel}
\end{equation}
The energy and momentum of an eigenstate are also given by simple functions of the rapidities,
namely
\begin{eqnarray}
E (\{ \lambda \}_N ) = \sum_{j} (\lambda_j^2 - h) \equiv E_{\lambda}, \hspace{1cm} P (\{ \lambda \}_N) = \sum_j \lambda_j \equiv P_{\lambda}
\end{eqnarray}
where $h$ is a chemical potential used to set the filling in a grand-canonical ensemble.

For our purposes, we also need explicit expressions for the form factors of the
field operators $\Psi (x, t)$ and $\Psi^{\dagger} (x,t)$ between Bethe eigenstates.
By making use of the fact that the energy and momentum are diagonal operators in this basis,
we can explicitly take the space and time dependence into account and write 
\begin{eqnarray}
\fl
\langle \{ \mu \}_{N-1} | \Psi (x, t) | \{ \lambda \}_{N} \rangle = 
e^{i (E_{\mu} - E_{\lambda}) t - i (P_{\mu} - P_{\lambda}) x} F (\{ \mu \}_{N-1}; \{ \lambda \}_N), \nonumber \\
\fl
\langle \{ \lambda \}_N | \Psi^{\dagger} (x, t) | \{ \mu \}_{N-1} \rangle = 
e^{i (E_{\lambda} - E_{\mu}) t - i (P_{\lambda} - P_{\mu}) x} {\bar F} (\{ \mu \}_{N-1}; \{ \lambda \}_N).
\end{eqnarray}
where 
\begin{equation}
F (\{ \mu \}_{N-1}; \{ \lambda \}_N) = \langle 0 | \prod_{j=1}^{N-1} C(\mu_j) \Psi (0, 0) \prod_{j=1}^N B(\lambda_j) | 0 \rangle
\end{equation}
and ${\bar F}$ is the conjugate.  In these notations, the correlation function (\ref{G2}) becomes
\begin{equation}
G_2 (x,t) = \sum_{\{ \mu \}_{N-1}} e^{i d (x, t; \lambda, \mu)} \frac{|F(\{ \mu \}_{N-1}; \{ \lambda \}_N)|^2}
{|| \{ \lambda \}_N || ^2 || \{ \mu \}_{N-1} ||^2}
\label{G2sum}
\end{equation}
in which
\begin{equation}
d(x,t; \lambda, \mu) = (E_{\lambda} - E_{\mu}) t - i (P_{\lambda} - P_{\mu}) x.
\end{equation}

The explicit evaluation of the summation in (\ref{G2sum}) to obtain a closed-form expression for the
correlation function remains an open problem in the theory of integrable models.  We will here however
make use of the building blocks which were developed in this spirit.  Namely, 
in \cite{KojimaCMP188}, the following useful expression was obtained for the field operator form factor:
\begin{eqnarray}
\fl
F (\{ \mu \}_{N-1}; \{ \lambda \}_N) = -i \sqrt{c} \prod_{N-1 \geq j > k \geq 1} g(\mu_j, \mu_k)
\prod_{N \geq j > k \geq 1} g(\lambda_k, \lambda_j) \times \nonumber \\
\times \prod_{l=1}^N \prod_{j=1}^N h (\lambda_l, \lambda_j)
\prod_{j=1}^{N-1} d(\mu_j) \prod_{l=1}^N d(\lambda_l) {\cal M}i (\{ \lambda \})
\label{formfactor_old}
\end{eqnarray}
\begin{equation}
{\cal M}i (\{ \lambda \}) = \left( 1 + \frac{\partial}{\partial \alpha} \right) \mbox{det}_{N-1} (S_{jk} - \alpha S_{Nk})|_{\alpha = 0}
\label{Mi_old}
\end{equation}
\begin{equation}
S_{jk} = t(\mu_k, \lambda_j) \frac{\prod_{a=1}^{N-1} h (\mu_a, \lambda_j)}{\prod_{a=1}^N h (\lambda_a, \lambda_j)}
- t(\lambda_j, \mu_k) \frac{\prod_{a=1}^{N-1} h (\lambda_j, \mu_a)}{\prod_{a=1}^N h (\lambda_j, \lambda_a)}.
\end{equation}
Although equation (\ref{formfactor_old}) offers an exact representation of the
form factors for fixed $N$, (\ref{Mi_old}) is not convenient for numerical purposes.
We therefore here first rewrite the form factors in terms of a single matrix determinant without derivative, which is much
more economically evaluated numerically.
The trick is simple, and consists in getting rid of the auxiliary parameter $\alpha$ in (\ref{Mi_old}) by using the
fact that, given two matrices $A$ and $B$ of which $B$ has rank $1$, the following identity holds:
\begin{equation}
\mbox{det} (A + \alpha B) = (1-\alpha) \mbox{det} A + \alpha \mbox{det} (A + B).
\end{equation}
Therefore,
\begin{equation}
\fl
{\cal M}i (\{ \lambda\}) = 
\left( 1 + \frac{\partial}{\partial \alpha} \right) \mbox{det}_{N-1} (S_{jk} - \alpha S_{N, k})|_{\alpha = 0}
= \mbox{det}_{N-1} (S_{jk} - S_{Nk}).
\label{Mi_one}
\end{equation}
Consider now the $N-1$ $\times$ $N-1$ matrix $W$ with entries
\begin{equation}
W_{jk} = \frac{1}{\mu_j - \lambda_k} \frac{\prod_{a=1}^N (\mu_j - \lambda_a)}{\prod_{a=1,\neq j}^{N-1} (\mu_j - \mu_a)}.
\end{equation}
Since the lines of this matrix are proportional to $1 \over (\mu_j - \lambda_k)$, the determinant of this
matrix is easily computed:
\begin{equation}
\mbox{det} W = \prod_{a > b=1}^{N-1} \frac{\lambda_b - \lambda_a}{\mu_b - \mu_a} 
\prod_{a=1}^{N-1} (\mu_a - \lambda_N).
\end{equation}
Consider now rewriting (\ref{Mi_one}) as
\begin{equation}
\mbox{det}_{N-1} (S_{jk} - S_{Nk}) = \frac{\mbox{det} \left( \sum_{l=1}^{N-1} (S_{jl} - S_{Nl}) W_{lk} \right)}{\mbox{det} W}.
\end{equation}
The matrix product above can be calculated explicitly.  Consider the sum
\begin{equation}
G_{jk}^+ = \sum_{l=1}^{N-1} t(\mu_l, \lambda_j) W_{lk}
\end{equation}
and the auxiliary contour integral
\begin{eqnarray}
I = \frac{1}{2\pi i} \oint_{|z| = R \rightarrow \infty} \frac{dz (ic)^2}{(z - \lambda_j) (z - \lambda_k) (z - \lambda_j + ic)}
\frac{\prod_{a=1}^N (z - \lambda_a)}{\prod_{a=1}^{N-1} (z - \mu_a)}.
\end{eqnarray}
The integration contour is a big circle of radius $R$ containing all poles of the integrand.  The value of this
integral is then given by the residue at infinity, which equals zero since the integrand behaves like
$1 \over z^2$ at $z \rightarrow \infty$.  On the other hand, the sum of the residues at points
$z = \mu_a$, $a = 1, ..., N-1$ inside the contour gives exactly $G_{jk}^+$.  In addition, there is a pole at
$z = \lambda_j - ic$ and, for $j = k$, we also have a pole at $z = \lambda_j$.  Since the sum over all of these residues 
is zero, we obtain
\begin{eqnarray}
\fl G_{jk}^+ = -ic \delta_{jk} \frac{\prod_{a \neq j}^N (\lambda_j - \lambda_a)}{\prod_{a=1}^{N-1} (\lambda_j - \mu_a)}
+ \frac{ic}{\lambda_k - \lambda_j + ic} \frac{\prod_{a=1}^N (\lambda_a - \lambda_j + ic)}{\prod_{a=1}^{N-1} (\mu_a - \lambda_j + ic)}.
\end{eqnarray}
Similarly, we have the identity
\begin{eqnarray}
G_{jk}^- = \sum_{l=1}^{N-1} t(\lambda_j, \mu_l) W_{lk} \nonumber \\
= ic \delta_{jk} \frac{\prod_{a \neq j}^N (\lambda_j - \lambda_a)}
{\prod_{a=1}^{N-1} (\lambda_j - \mu_a)}
- \frac{ic}{\lambda_j - \lambda_k + ic} \frac{\prod_{a=1}^N (\lambda_j - \lambda_a + ic)}{\prod_{a=1}^{N-1} (\lambda_j - \mu_a + ic).}
\end{eqnarray}
This whole procedure allows us to rewrite the matrix product 
$\sum_{l=1}^{N-1} (S_{jl} - S_{Nl}) W_{lk}$, finally leading to
\begin{eqnarray}
\fl {\cal M}i (\{ \lambda \}) = 
i^{N-1} c^{2(N-1)} \frac{\prod_{a > b}^{N} (\lambda_a - \lambda_b) \prod_{a > b}^{N-1} (\mu_b - \mu_a)}
{\prod_{a=1}^N \prod_{b=1}^{N-1} (\lambda_a - \mu_b)} \mbox{det}_{N-1} U_{jk},
\end{eqnarray}
in which the matrix $U$'s entries are functions of the rapidities of the left and right eigenstates,
\begin{eqnarray}
\fl U_{jk} (\{ \mu \}, \{ \lambda \}) = \delta_{jk} (\frac{V^+_j - V^-_j}{i}) + \frac{\prod_{a=1}^{N-1} (\mu_a - \lambda_j)}
{\prod_{a\neq j}^{N} (\lambda_a - \lambda_j)} (K(\lambda_j - \lambda_k) - K (\lambda_{N} - \lambda_k))
\label{U}
\end{eqnarray}
in which
\begin{equation}
V^{\pm}_j = \frac{\prod_{a=1}^{N-1} (\mu_a - \lambda_j \pm ic)}{\prod_{a=1}^{N} (\lambda_a - \lambda_j \pm ic)}.
\end{equation}
For the repulsive Bose gas which we are considering, the $U$ matrix has purely real entries.
The matrix elements also conveniently coincide with those obtained in \cite{SlavnovTMP79_82} for the
(integral of the) local density operator, and which we used in \cite{CauxPRA74} to compute
density correlation functions.

\section{Dynamical one-particle correlation function}

From the results of the previous section, we can therefore rewrite the one-body dynamical correlation
function (\ref{G2}) as
\begin{eqnarray}
G_2 (x,t) \equiv \langle \Psi^{\dagger} (x, t) \Psi (0, 0) \rangle_N = \sum_{\{ \mu \}_{N-1}} e^{i d(x,t; \lambda, \mu)} 
G (\{ \mu \}, \{\lambda \}), 
\end{eqnarray}
in which the correlation weight for a given intermediate state is explicitly given by
the following function of its rapidities $\{ \mu \}_{N-1}$ and of the set of ground state rapidities $\{ \lambda \}_N$:
\begin{eqnarray}
\fl G (\{ \mu \}, \{ \lambda \}) = 
c^{2N - 1} \frac{\prod_{j>k=1}^N (\lambda_{jk}^2 + c^2)^2}{\prod_{a=1}^N \prod_{b=1}^{N-1} (\lambda_a - \mu_b)^2} 
\frac{[\mbox{det}_{N-1} U (\{ \mu \}, \{ \lambda \})]^2}{||\{ \lambda \}_N ||^2 ||\{\mu \}_{N-1}||^2}
\end{eqnarray}
where the $U$ matrix entries are given by equation (\ref{U}),
and the state norms are given by (\ref{norm}) and (\ref{Gaudin}).

In the present context, it is more convenient to consider the space and time Fourier transform of the correlation function, 
\begin{eqnarray} 
G_2 (k, \omega) = \int_0^L dx \int_{-\infty}^{\infty} dt e^{i \omega t - i k x} G_2 (x,t) \nonumber \\
= 2\pi L \sum_{\{ \mu \}_{N-1}} \delta(\omega - E_{\{\mu\}} + E_{\{ \lambda \}}) \delta_{k, P_{\{ \mu \}} - P_{\{ \lambda \}}}
G (\{ \mu\}, \{ \lambda \}),
\label{G2komega}
\end{eqnarray}
which explicitly displays that each intermediate Bethe eigenstate contributes all its correlation weight
$G$ at a single point in the $k$, $\omega$ plane.  

A particular intermediate state can be composed of
a number of single-particle excitations.  Due to the fermion-like structure of
the ground state, these take two different basic
forms, namely (following the terminology of \cite{LiebPR130}) Type I (particle) and Type II (hole).
Particle-like excitations are obtained by adding an extra occupied quantum number outside the ground state interval,
while the hole-like excitations are obtained by removing one of the occupied ground state quantum numbers.
Type I modes are Bogoliubov-like particles existing at any momentum, and whose dispersion relation 
(in the infinite system) interpolates between $\varepsilon_I (k) = k^2$ at the noninteracting $\gamma = 0$ 
point to $\varepsilon_I(k) = k^2 + 2\pi n|k|$ (where $n=N/L$ is the density) as $\gamma \rightarrow \infty$.  Type II particles do not
appear in Bogoliubov theory.  Their dispersion relation coincides with the lower threshold of the 
correlation function, and vanishes at $k = 2k_F$, with the Fermi momentum given by $k_F = \pi n$.
Both types of excitations have a dispersion relation which approaches $k \rightarrow 0$ with a slope equal to the velocity of sound $v_s (\gamma)$.

The action of the field operator
on a Bethe eigenstate is found to be relatively innocuous, in the sense that the size distribution of its matrix 
elements in this basis is very strongly peaked.  There are in other words only relatively few form factors of
large value, and all of these involve only up to a handful of elementary excitations.  
The trace over the Fock space of intermediate states can thus be performed in a highly optimized way
by searching for and concentrating the computational effort on these important intermediate states
(part of the idea of the ABACUS method).  This is implemented with a recursive search algorithm.

In order to verify the accuracy of our results, we make use of the observation that
integrating the one-body correlation function (\ref{G2komega}) over frequency and summing over momenta simply yields
back the density of the gas,
\begin{eqnarray}
\int_{-\infty}^{\infty} \frac{d\omega}{2\pi} \frac{1}{L} \sum_k G_2 (k, \omega) = \sum_{\{ \mu \}_{N-1}} G (\{\mu \}, \{ \lambda \})
\nonumber \\
= G_2 (x=0, t = 0) = \langle \Psi^{\dagger} (0,0) \Psi(0,0) \rangle = n,
\label{sumrule}
\end{eqnarray}
which provides a convenient sumrule for our method.  Physically, for large enough $N$ and $L$, there is only one
important physical parameter, which is defined as $\gamma = c/n$.  All the results presented here are
obtained by setting the density to one, and varying the interaction parameter $c$.  

\begin{figure}
\begin{tabular}{ll}
\includegraphics[width=7cm]{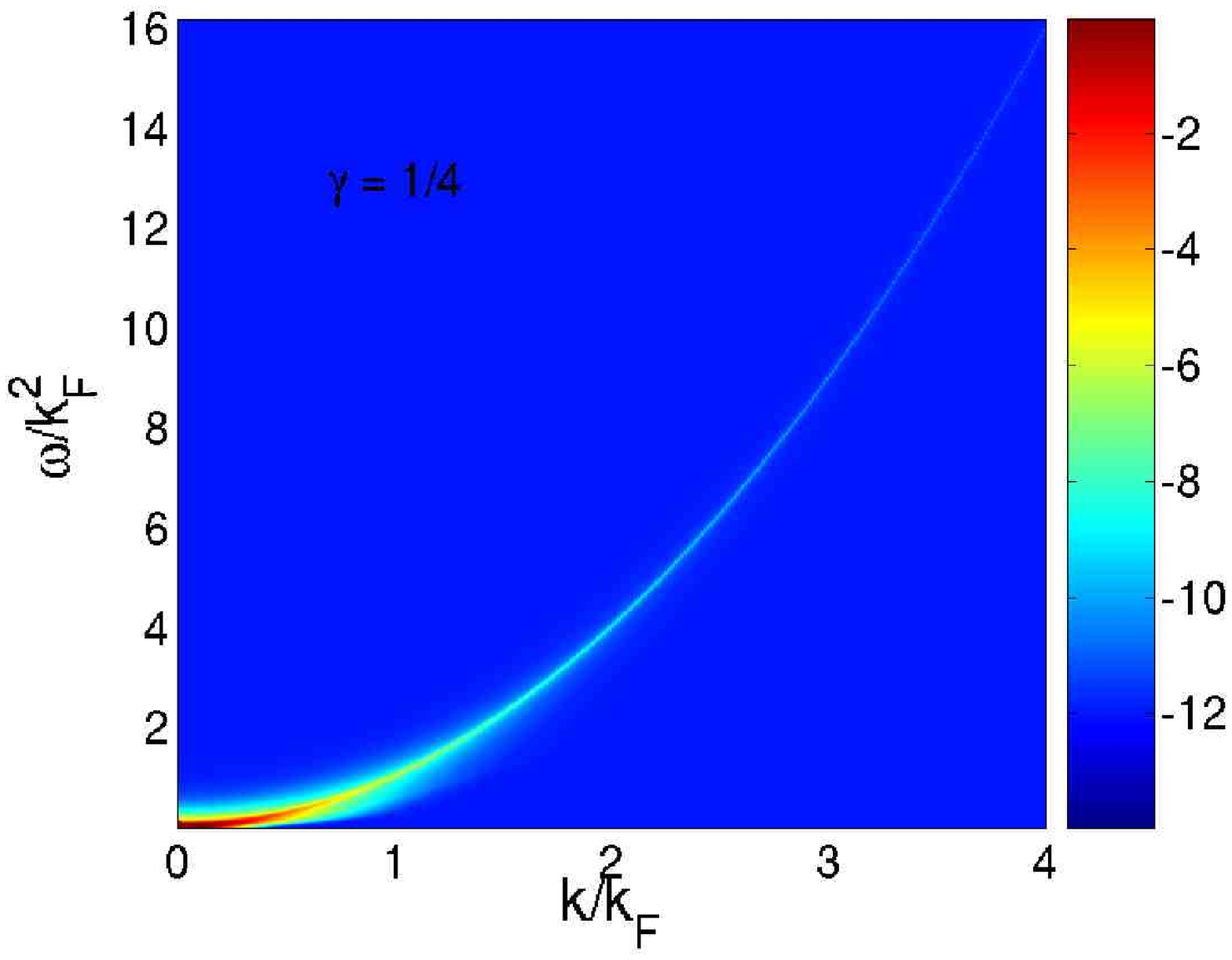} &
\includegraphics[width=7cm]{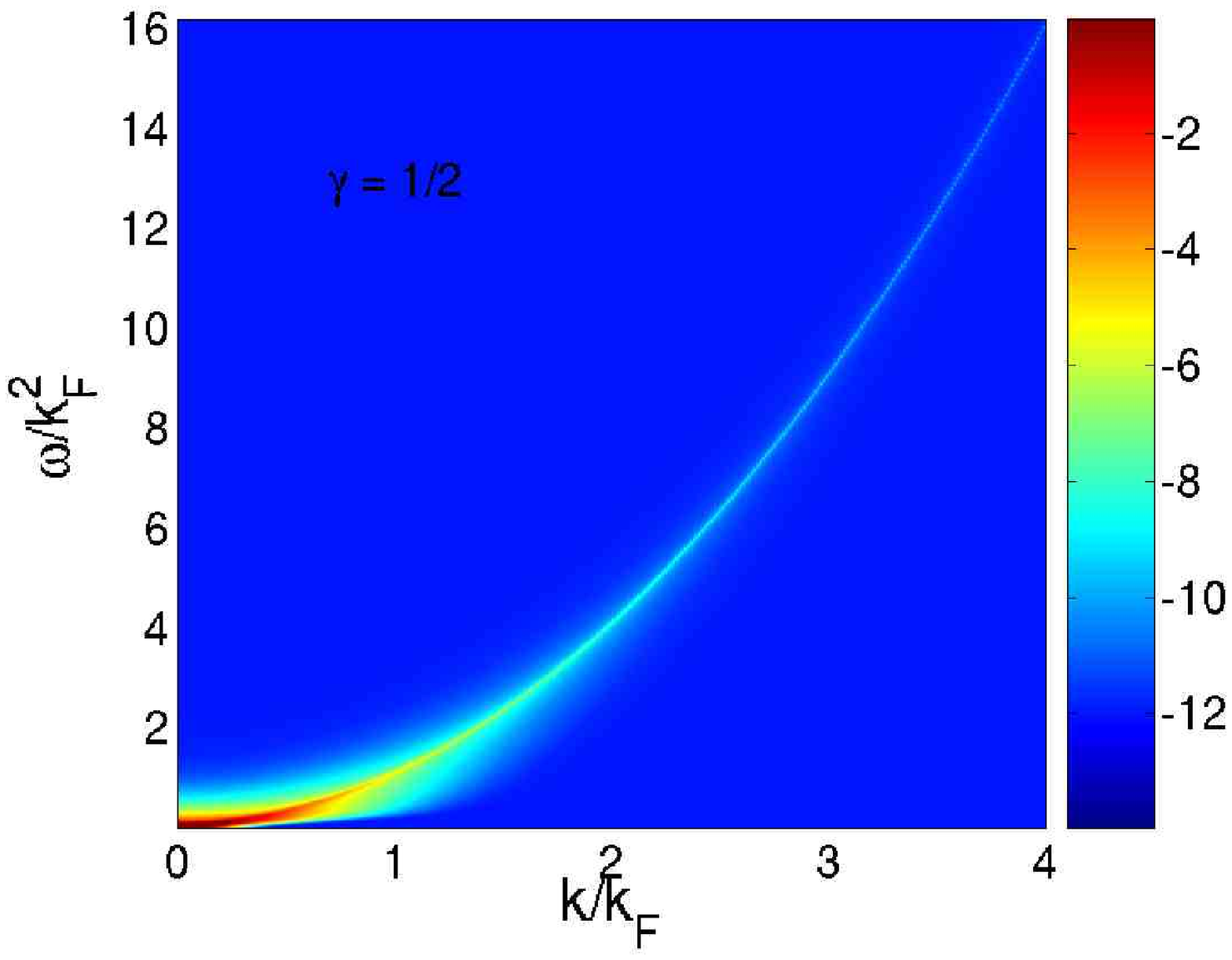} \\
\includegraphics[width=7cm]{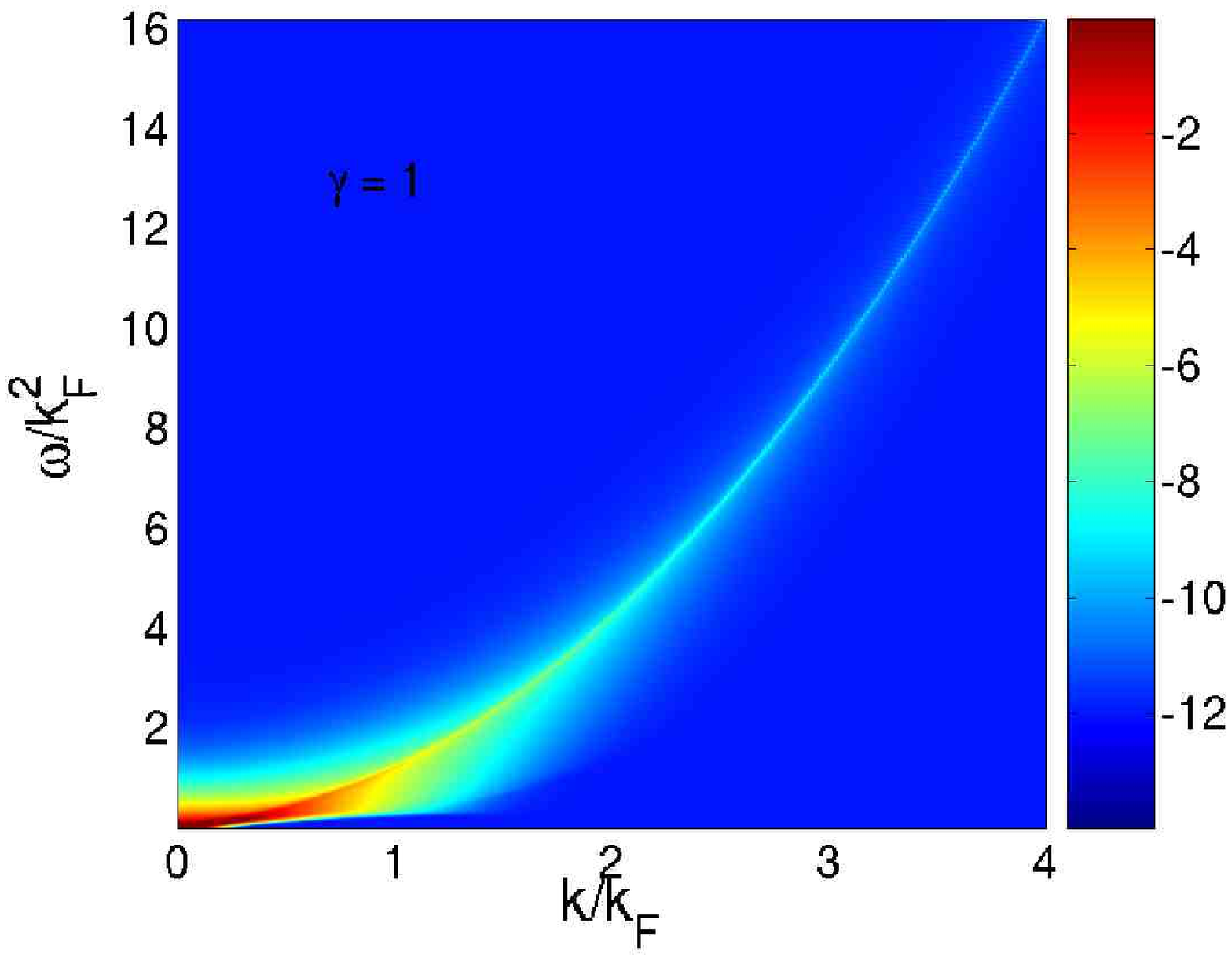} &
\includegraphics[width=7cm]{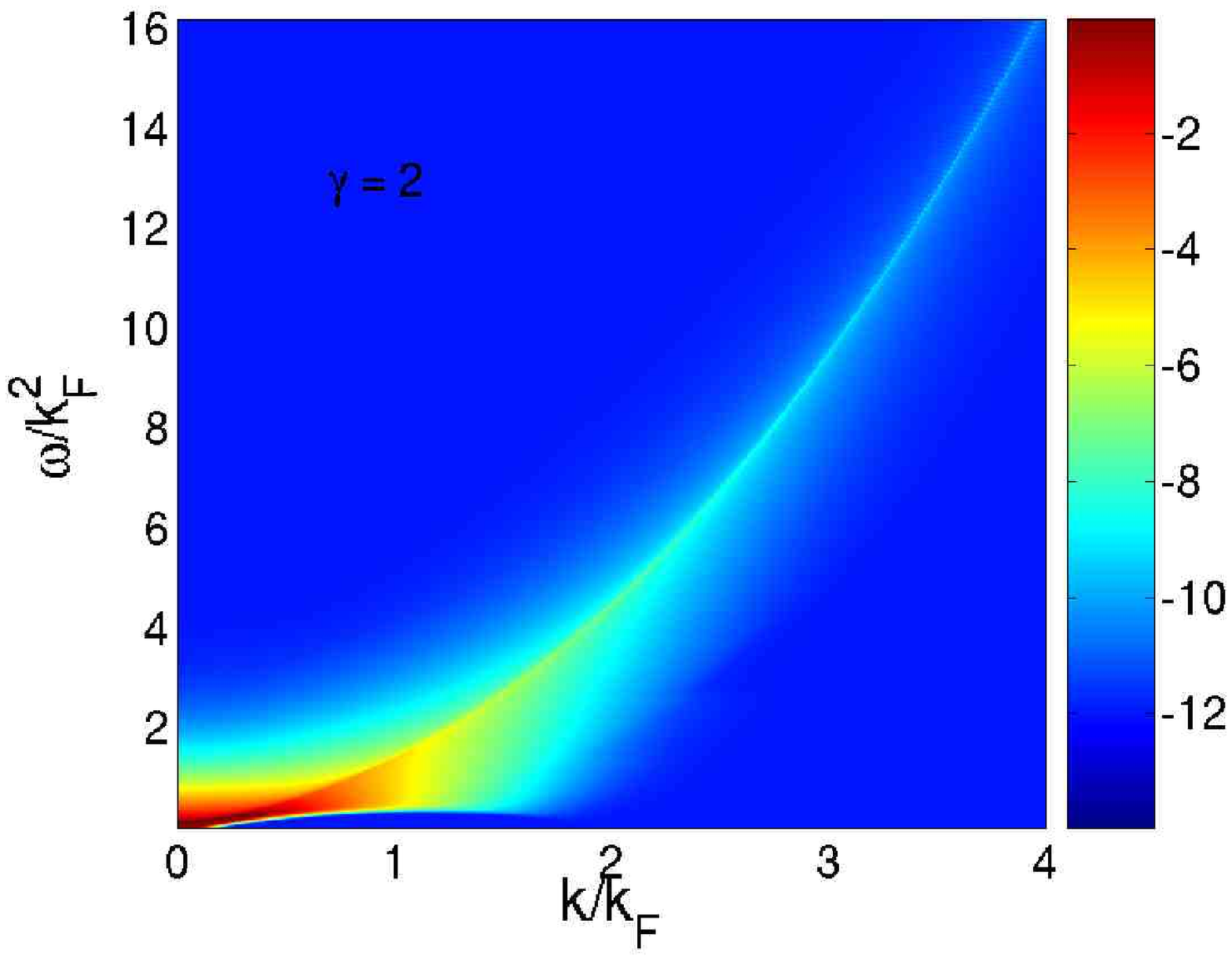} \\
\includegraphics[width=7cm]{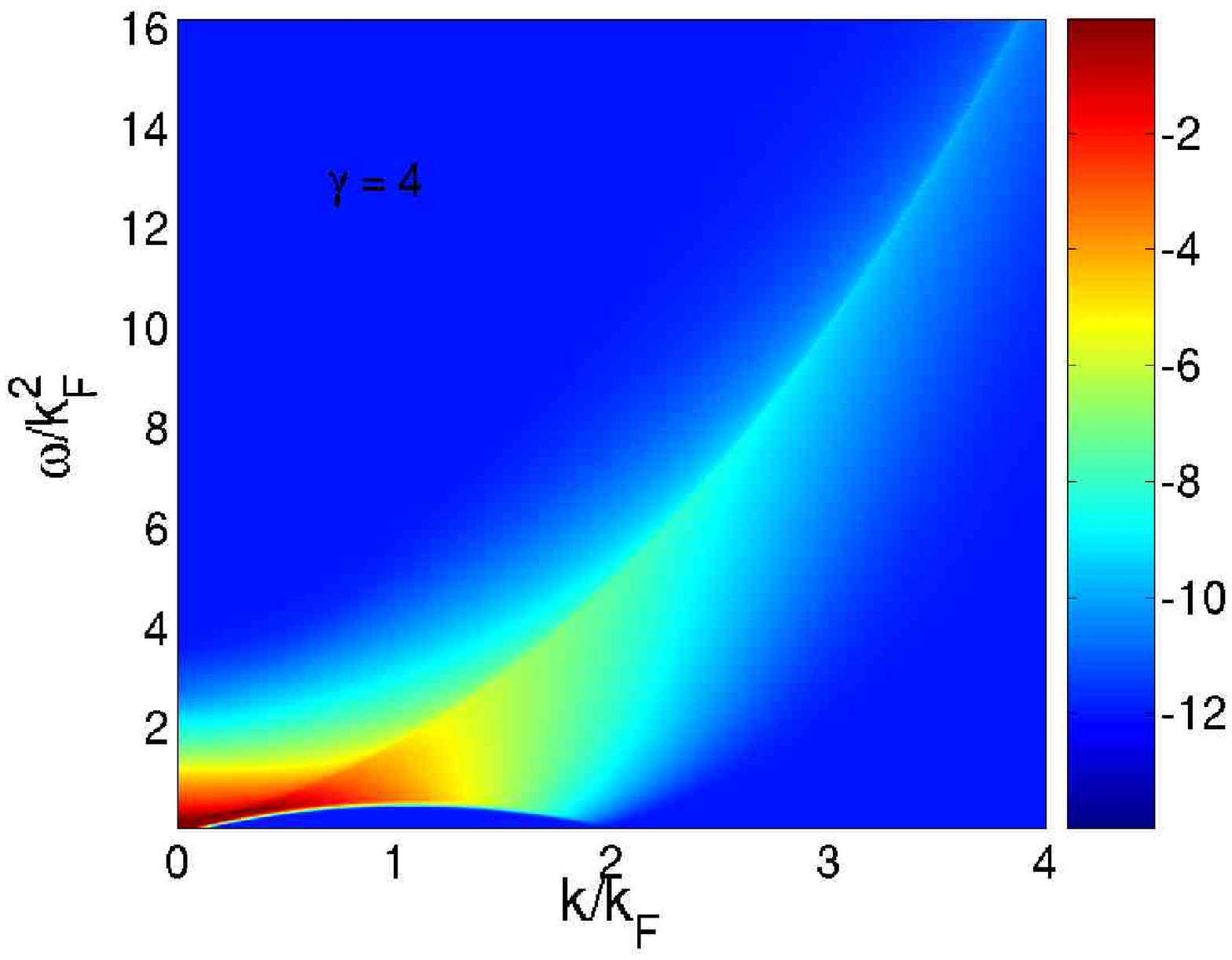} &
\includegraphics[width=7cm]{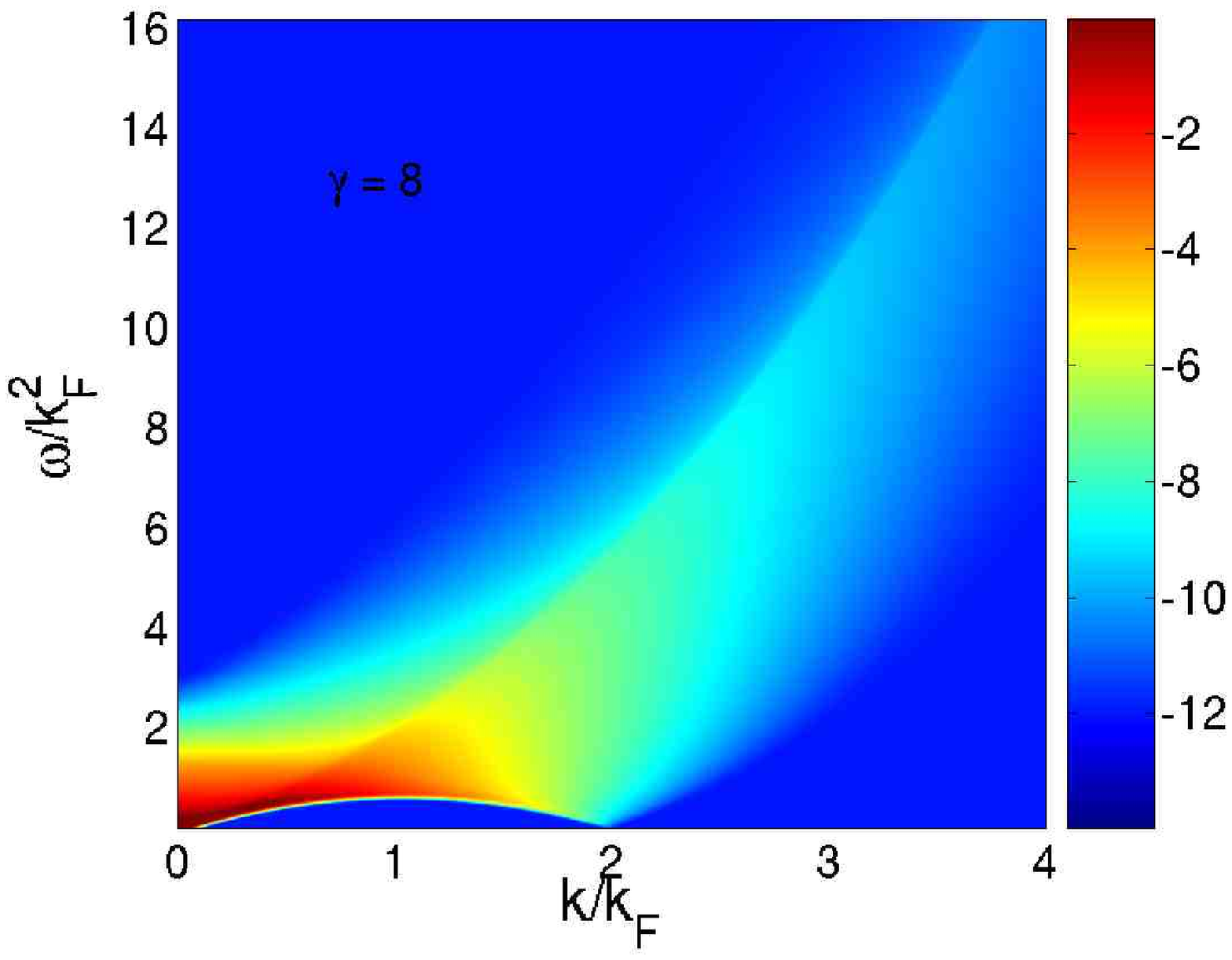} 
\end{tabular}
\caption{Pseudocolor plots of the logarithm of the dynamical one-body correlation function of the Lieb-Liniger gas, for $\gamma = 1/4, 1/2, 1, 2, 4$ and $8$.  
The data was obtained from systems at unit density and $N = 150$ particles.  The horizontal axis is momentum,
and the vertical one energy.  Sum rule saturations are listed in Table \ref{SRtable}.  At small $\gamma$, the correlation
weight at fixed momentum falls mostly around the Type I dispersion relation.}
\label{GF1}
\end{figure}
\begin{figure}
\begin{tabular}{ll}
\includegraphics[width=7cm]{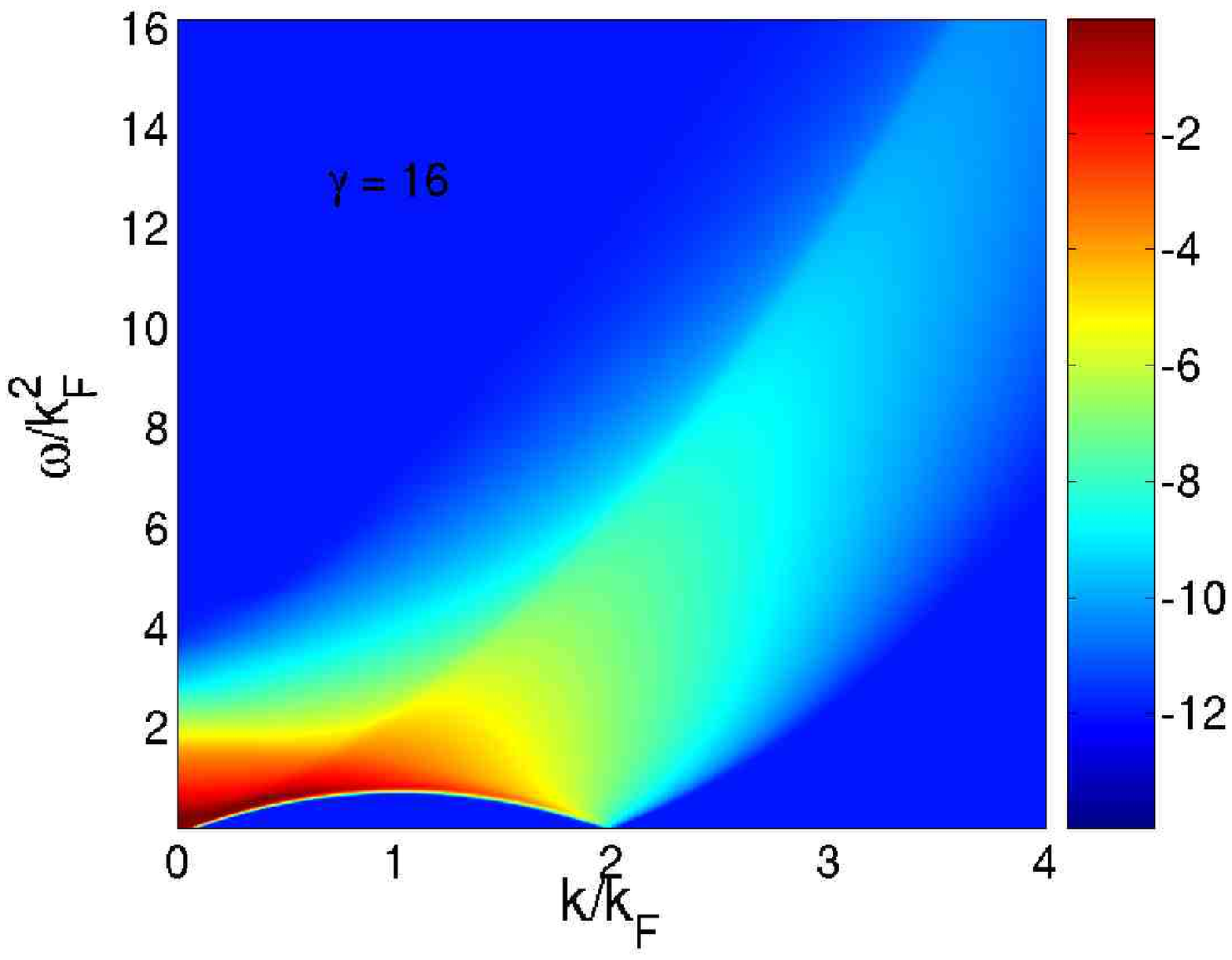} &
\includegraphics[width=7cm]{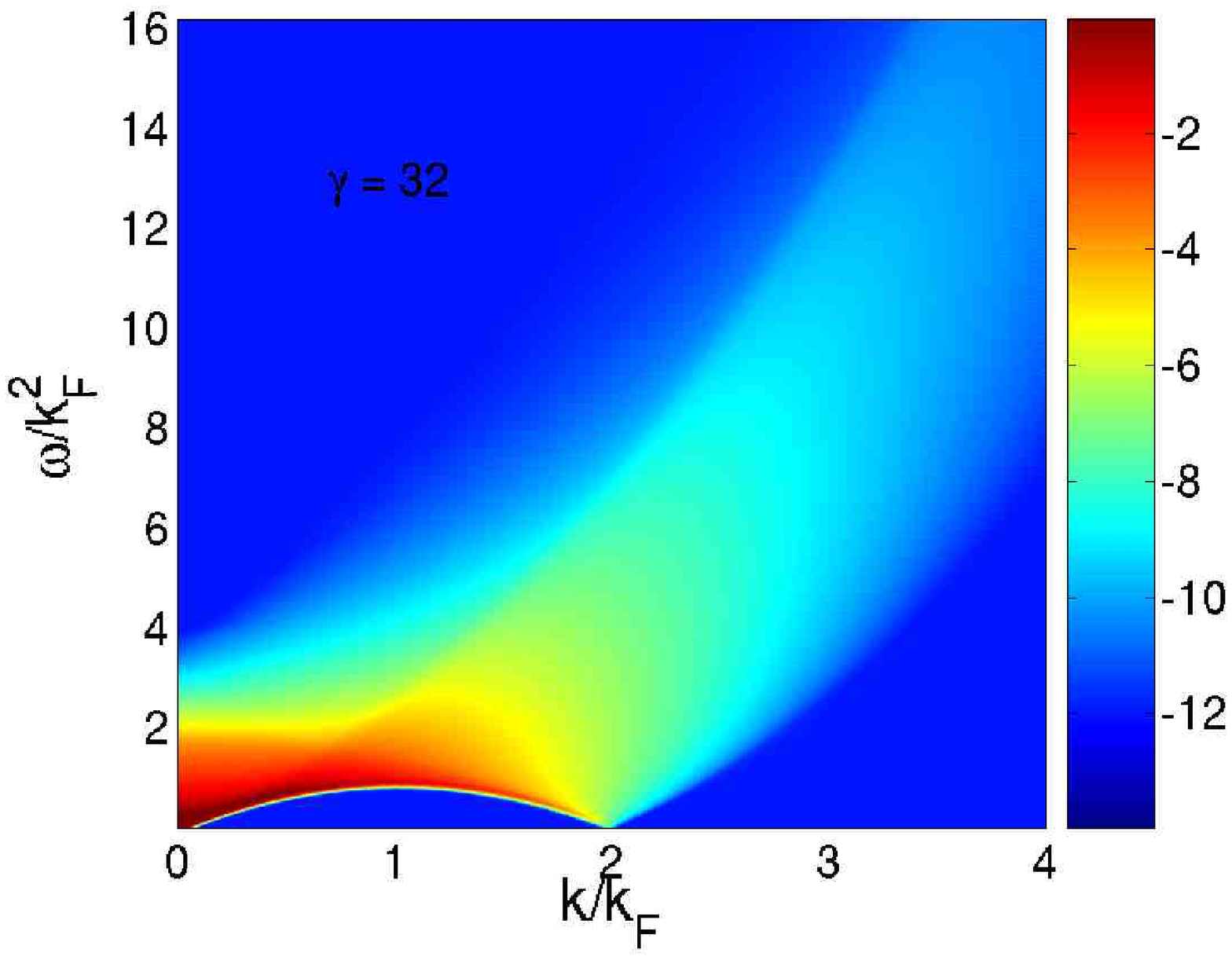} \\
\includegraphics[width=7cm]{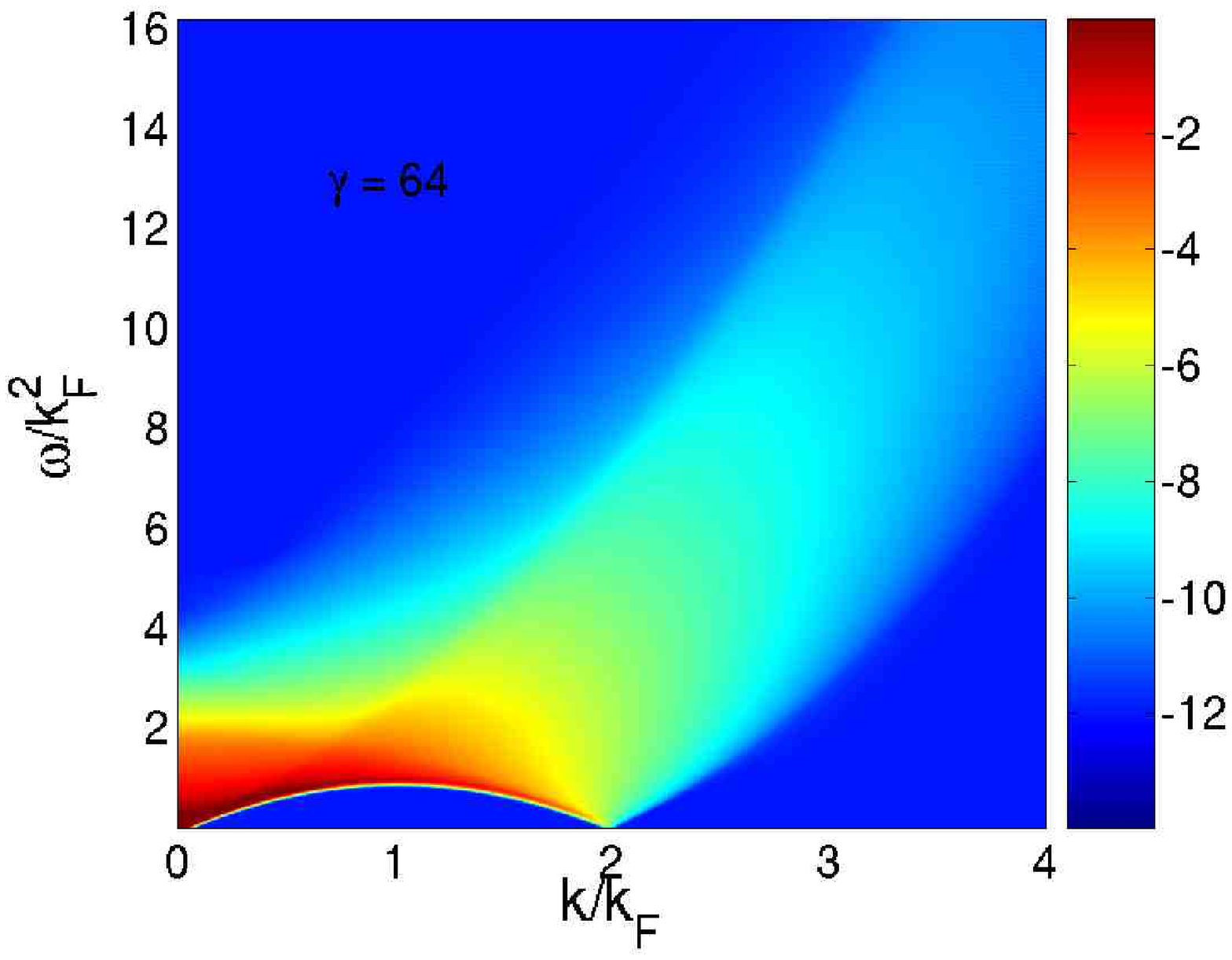} &
\includegraphics[width=7cm]{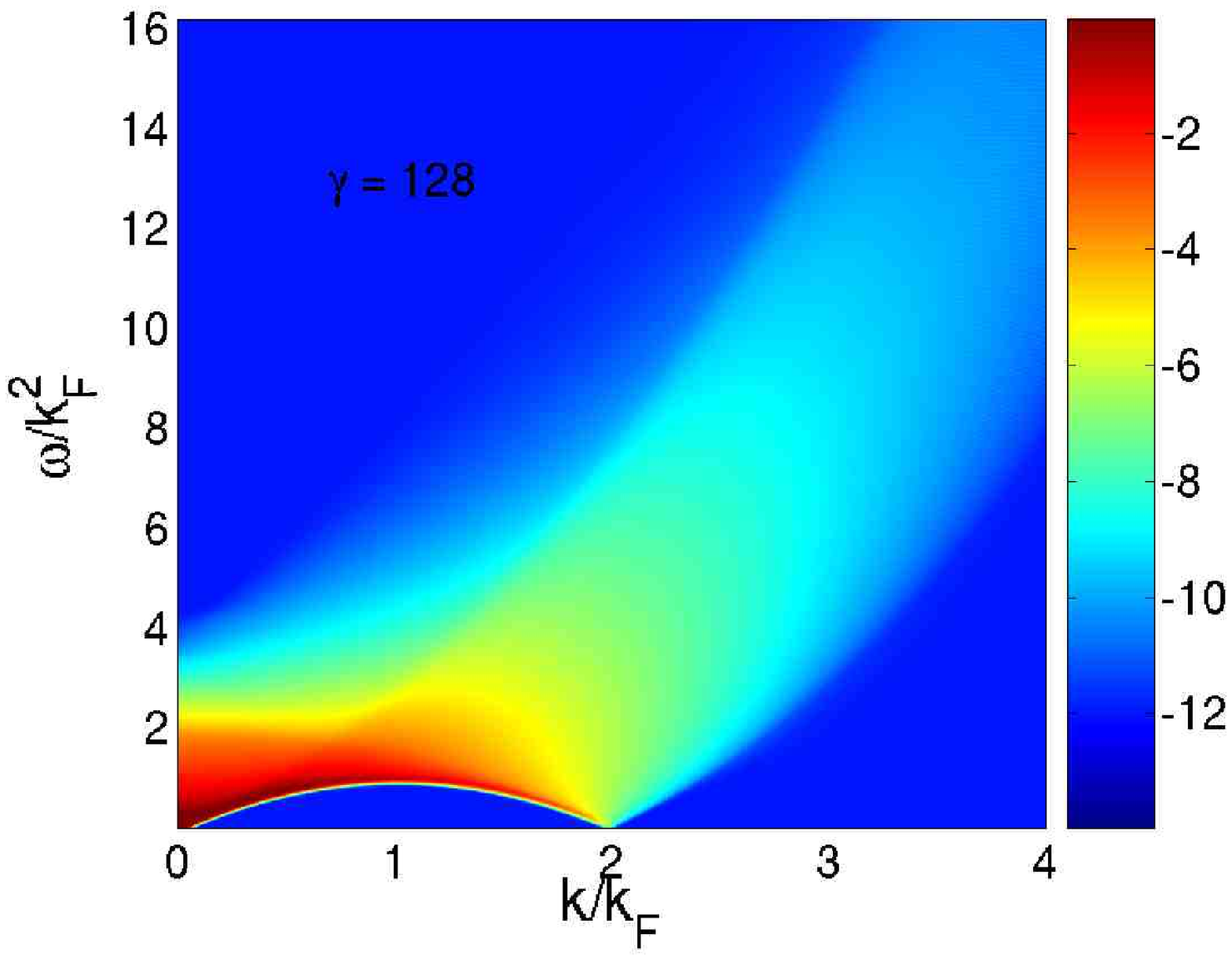} 
\end{tabular}
\caption{Same as Figure \ref{GF1}, but now for $\gamma = 16, 32, 64$ and $128$, showing the progression of the distribution of
correlation weight in the strongly-interacting regime.  As $\gamma$ increases, more correlation weight gets
concentrated near the lower threshold.}
\label{GF2}
\end{figure}

\subsection{Results}

\begin{table}[b]
\caption{Sum rule saturation percentages achieved as a function of the interaction
parameter, for various values of $N$.  The last line gives the contribution coming from states outside of
the sub-Fock space used for the numerics, as computed from the asymptotic form of the static correlator (see text).}
\begin{tabular}{|c|c|c|c|c|c|c|c|c|c|c|}
\hline
$N,  \gamma$ & 1/4 & 1/2 & 1 & 2 & 4 & 8 & 16 & 32 & 64 & 128 \\ \hline \hline
$50$ & 100 & 100 & 100 & 99.99 & 99.98 & 99.96 & 99.94 & 99.92 & 99.91 & 99.91 \\ \hline
$100$ & 100 & 100 & 99.99 & 99.98 & 99.96 & 99.92 & 99.85 & 99.81 & 99.78 & 99.75 \\ \hline
$150$ & 100 & 99.99 & 99.98 & 99.95 & 99.91 & 99.71 & 99.61 & 99.33 & 99.44 & 99.20 \\ \hline
$200$ & 99.99 & 99.96 & 99.92 & 99.87 & 99.64 & 99.28 & 98.83 & 98.59 & 98.43
& 98.33 \\ 
\hline
\hline
$k>4k_F$& $2\cdot 10^{-4}$&$8\cdot 10^{-4}$ & 0.003 & 0.007 & 0.018 & 0.032 & 0.047 & 0.058 & 0.064 & 0.067\\
\hline
\end{tabular}
\label{SRtable}
\end{table}

Our results for the dynamical one-body correlation function (\ref{G2}) for
different values of $\gamma$ are presented in Figures \ref{GF1} and \ref{GF2}
as density plots.
To obtain continuous curves, the energy delta functions in (\ref{G2komega})
are smoothened into Gaussians of width
slightly larger than the typical interlevel spacing.  The color coding of
these  figures follows the logarithm of the correlation function.  
All data presented in these was obtained by considering systems of $N = 150$
particles with unit density.
For each individual plot, around seventy million intermediate states were
taken into account, yielding extremely good saturation of the sum 
rule (\ref{sumrule}), as summarized in Table \ref{SRtable}.  
The summation we actually perform is restricted to intermediate states within
a window of momentum $k \in [0, 4k_F]$, composed of (possibly many) single-particle
excitations of momentum $|k| < k_{max} = 8 k_F$.  The correlation weight carried by
states outside of this sub-Fock space is negligible, as can be clearly seen from
the sum rule saturations achieved.  This can also be independently checked 
because the large $k$ tail of the
static correlation function is exactly known from \cite{od-03} (as discussed 
more extensively in the next section, see Eq. (\ref{k-4}) below). 
In the last row of Table \ref{SRtable} we report the contributions calculated
from Eq. (\ref{k-4}) for all the states with $k>4k_F$. 
We checked that the error due to finite $N$ and to subleading corrections to 
Eq. (\ref{k-4}) do not change the accuracy we reported.  
From these data it is evident that states with $k>4 k_F$ contribute 
for almost all the missing part for $N=50$ and for a relevant part for $N=100$.
In all cases, the (rest of the) missing part of
the sum rules is therefore ascribed to intermediate states within the sub-Fock
space we consider, but which we do not include in the summation.  In fact, our
sum only considers an extremely small fraction of all the available intermediate
states within the sub-Fock space, made of only a handful of elementary excitations.  The distribution of
correlation weight between $0, 2, 4, 6$ and $8$ particle states for a given value
$\gamma = 16$ for different values of $N$ is given in Table 2 (here, $2n$ particle states 
are made from $n$ particle-hole excitations on the ground state of $N-1$ particles).  
These numbers represent lower bounds of the contributions from the different classes of 
states.  There are large variations of this distribution as a function of $N$, although
the full correlator itself remains more or less invariant.  The general scenario is
that $0$ and $2$ particle contributions are strictly decreasing functions of $N$, 
while higher particle number contributions first increase and thereafter decrease, with a maximum
at $N^{max}_{2n}$ such that $N^{max}_{2n} > N^{max}_{2m}$ if $n > m$.  As a function of $\gamma$, 
we observe as naturally expected that for small $\gamma$, intermediate states with smaller (larger) numbers of particles carry more (less) weight,
with more weight shifting to higher particle numbers when $\gamma$ increases.

We also provide in the left panels of Figures 
3 and 4 
a series of fixed-momentum cuts for a number of values of the
interaction parameter, with data for two different system sizes, smoothened as explained above
(without smoothing, these would be $\delta$ peaks;  the smoothing however blurs the lower
threshold slightly).  At small $\gamma$, the 
correlation is peaked around the Type I dispersion relation.  As the interaction parameter increases,
the peak becomes broader and moves towards lower energies, as expected in view of the increasing
fermion-like nature of the system.  At intermediate values of $k$, the correlation
width also increases with $\gamma$, with the peaks eventually shifting altogether from high to low energies.
The difference between the $N = 100$ and $N = 150$ curves can mostly be ascribed to imperfect sumrule
saturation (the logarithm of the correlation function is plotted, and therefore the actual difference
between curves lies in regions where the correlation weight is extremely small).  

On the right panels of these figures, the same data sets are used to plot the integrated correlation,
\begin{eqnarray}
G^I_2 (k, \omega) = \int_0^{\omega} d\omega' G_2 (k, \omega) = \nonumber \\
= 2\pi L \sum_{\{ \mu \}_{N-1}} 
\Theta (\omega' - E_{\{ \mu \}} + E_{\{ \lambda \}}) \delta_{k, P_{\{ \mu \}} - P_{\{ \lambda \}}}
G (\{ \mu\}, \{ \lambda \}),
\label{GI}
\end{eqnarray}
for which no smoothing procedure is necessary.  These curves are in fact series of steps of height
corresponding to the correlation weight of any intermediate state lying at this point in $k$, $\omega$ space.
In the infinite size limit, these would become smooth curves whose $\omega$ derivative would be
the one-body correlation function.  The discreteness of the steps due to the
finite number of particles can just be made out in the plots.
The fact that no artificial smoothing of the delta functions is used in these data sets
makes them the most appropriate objects for eventual comparison with other theoretical or
numerical methods (as e.g. those employed in Refs. \cite{other} for the
dynamical structure factor).

\begin{table}
\caption{Distribution of sum rule weight between intermediate states involving $N_p$ particles, with $N_p = 0, 2, 4, 6$ and $8$.
The numbers given are the fraction of the total sumrule associated to each sub-type of excited state
which was obtained during the runs.  We present data for $\gamma = 16$, for three values of $N$.  These are lower
bounds only, since we only scan over a very small fraction of the total number of multiparticle states.}
\begin{tabular}{|c|c|c|c|c|c|c|}
\hline
$N, N_p$ & 0 & 2 & 4 & 6 & 8 & TT \\ \hline \hline
50 & 0.29075 & 0.47437 & 0.22541 & 0.00877 & 0.00006 & 0.99936\\ \hline 
100 & 0.22157 & 0.46575 & 0.29119 & 0.01979 & 0.00018 & 0.99848 \\ \hline
150 & 0.18881 & 0.45288 & 0.32627 & 0.02813 & --- & 0.99609 \\ \hline
\end{tabular}
\label{sr_mp}
\end{table}

\begin{figure}
\begin{tabular}{ll}
\includegraphics[width=7cm]{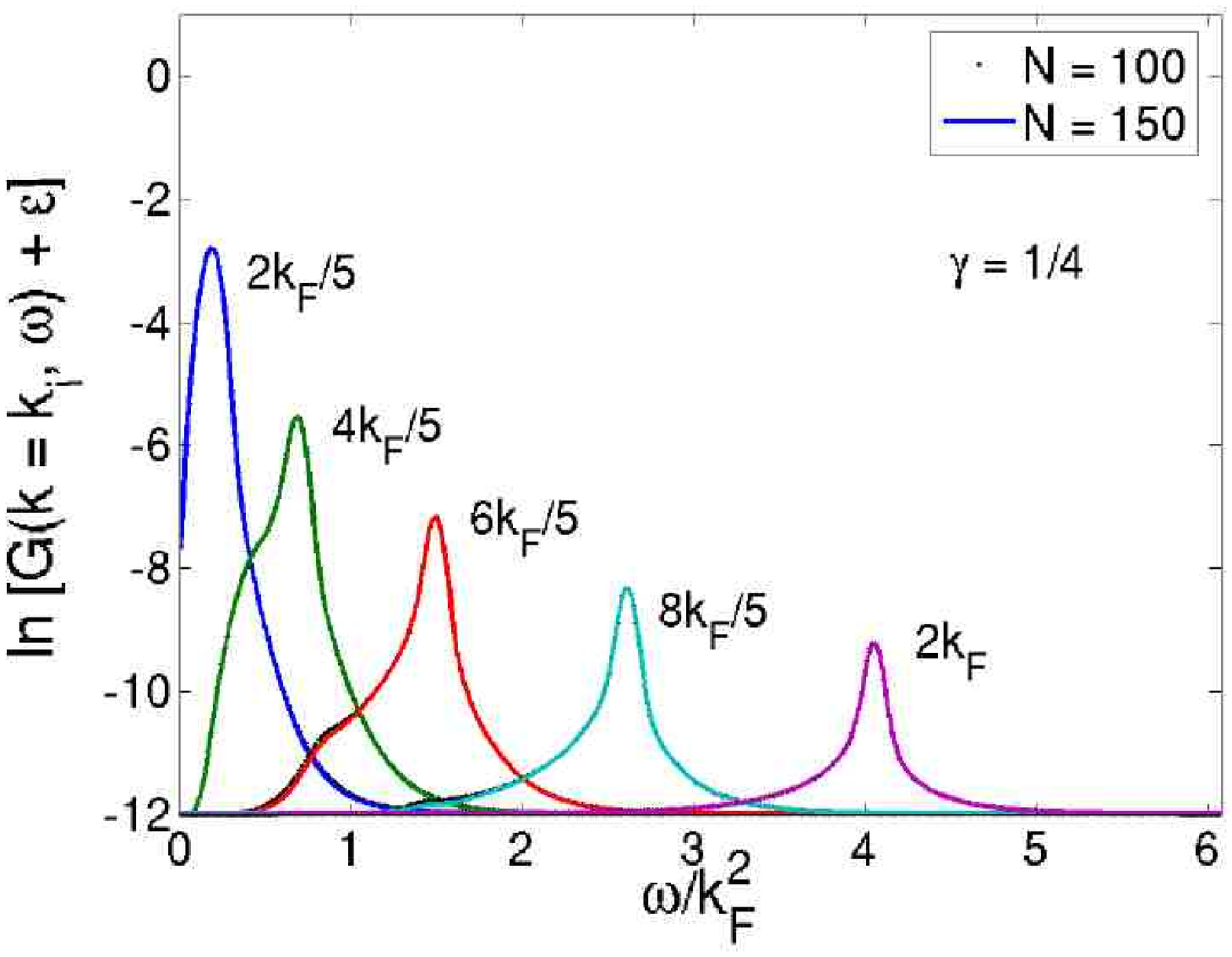} &
\includegraphics[width=7cm]{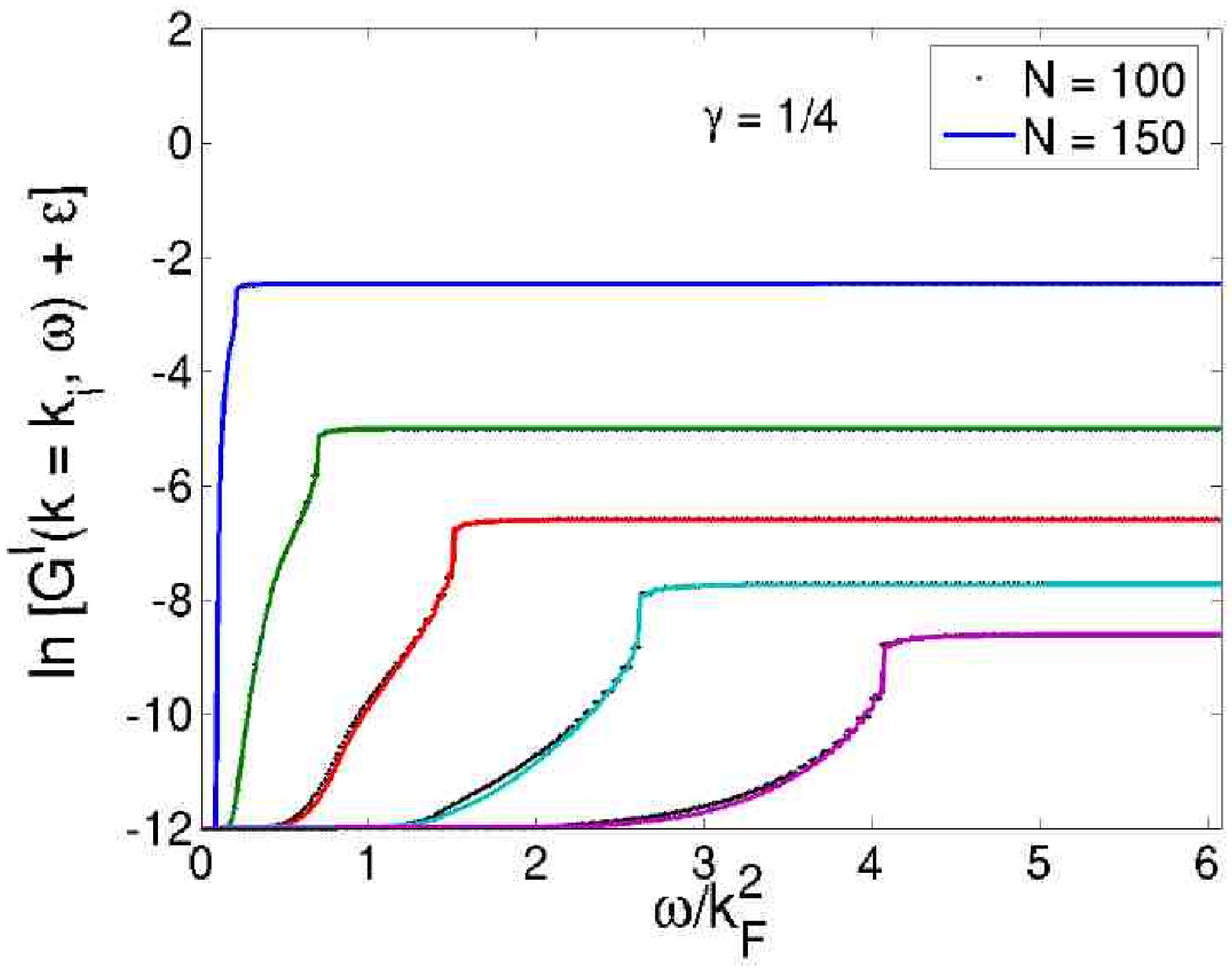} \\
\includegraphics[width=7cm]{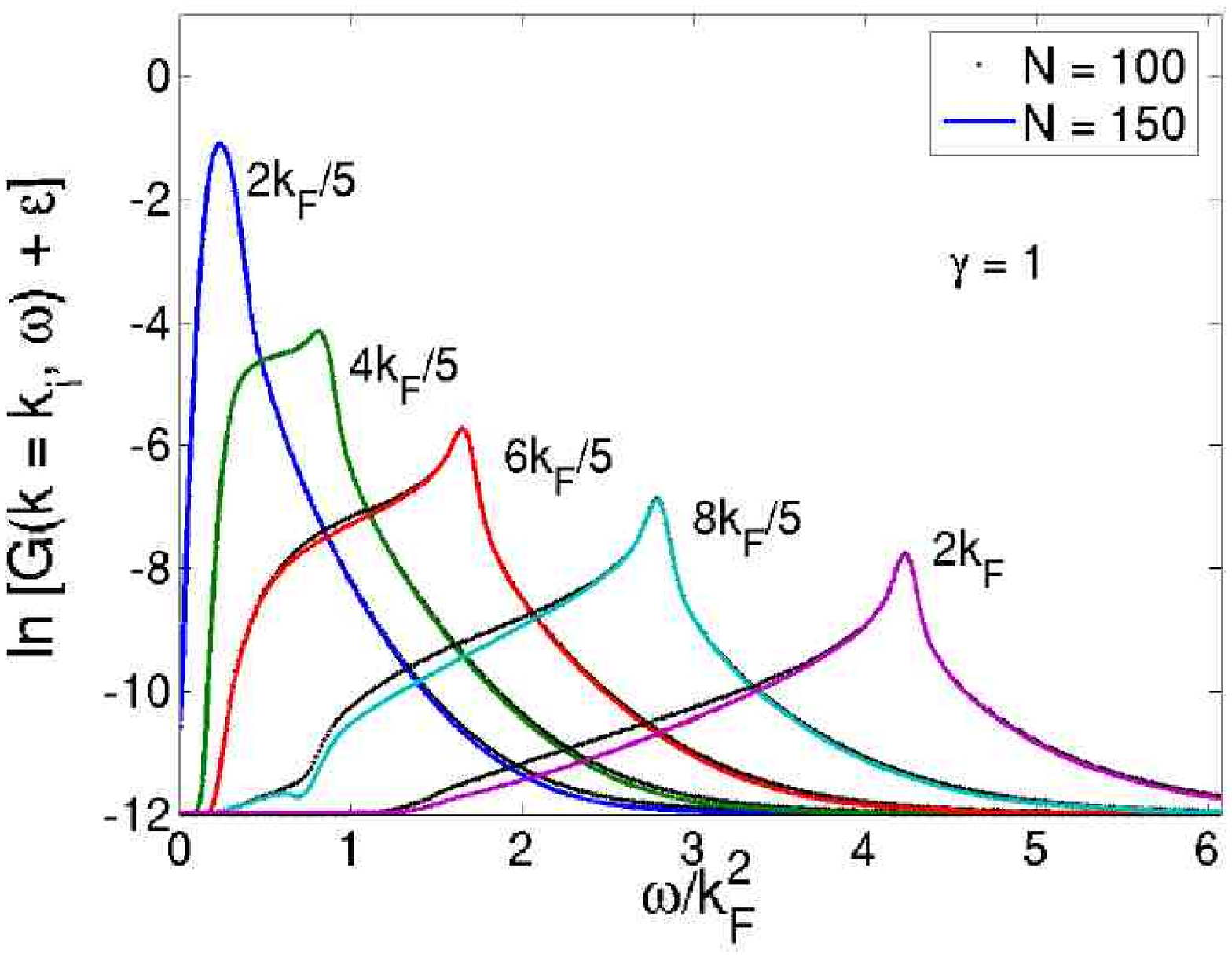} &
\includegraphics[width=7cm]{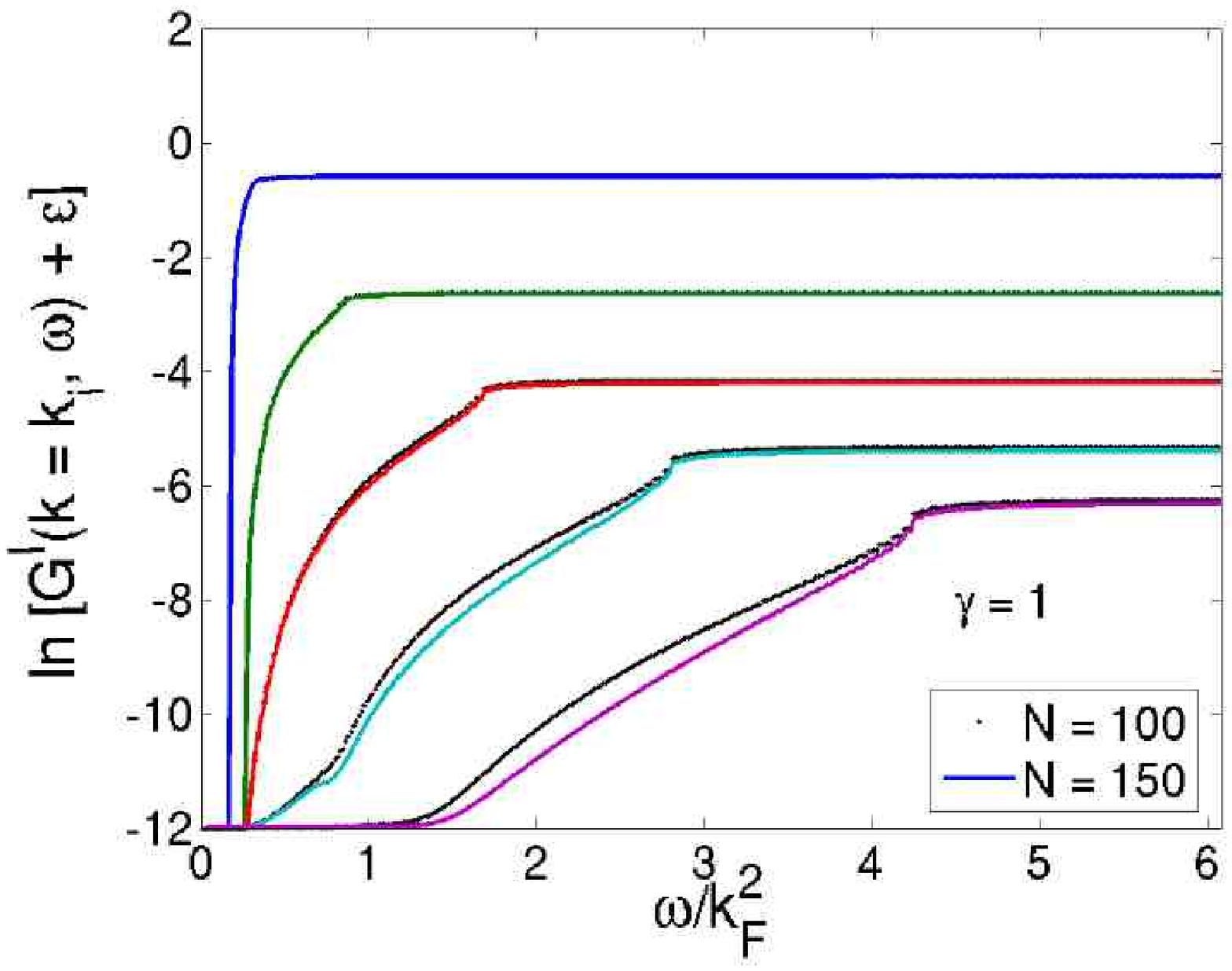} \\
\includegraphics[width=7cm]{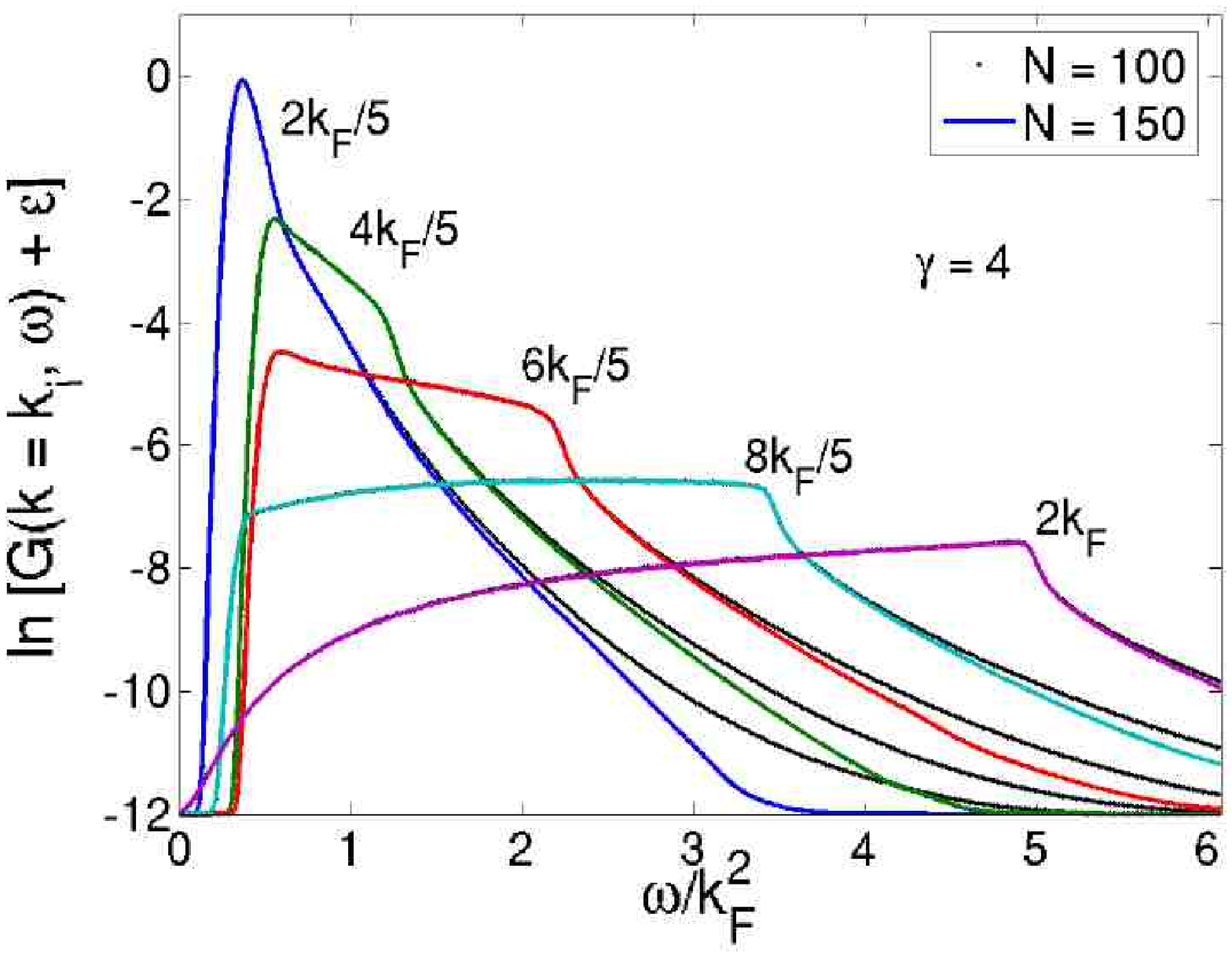} &
\includegraphics[width=7cm]{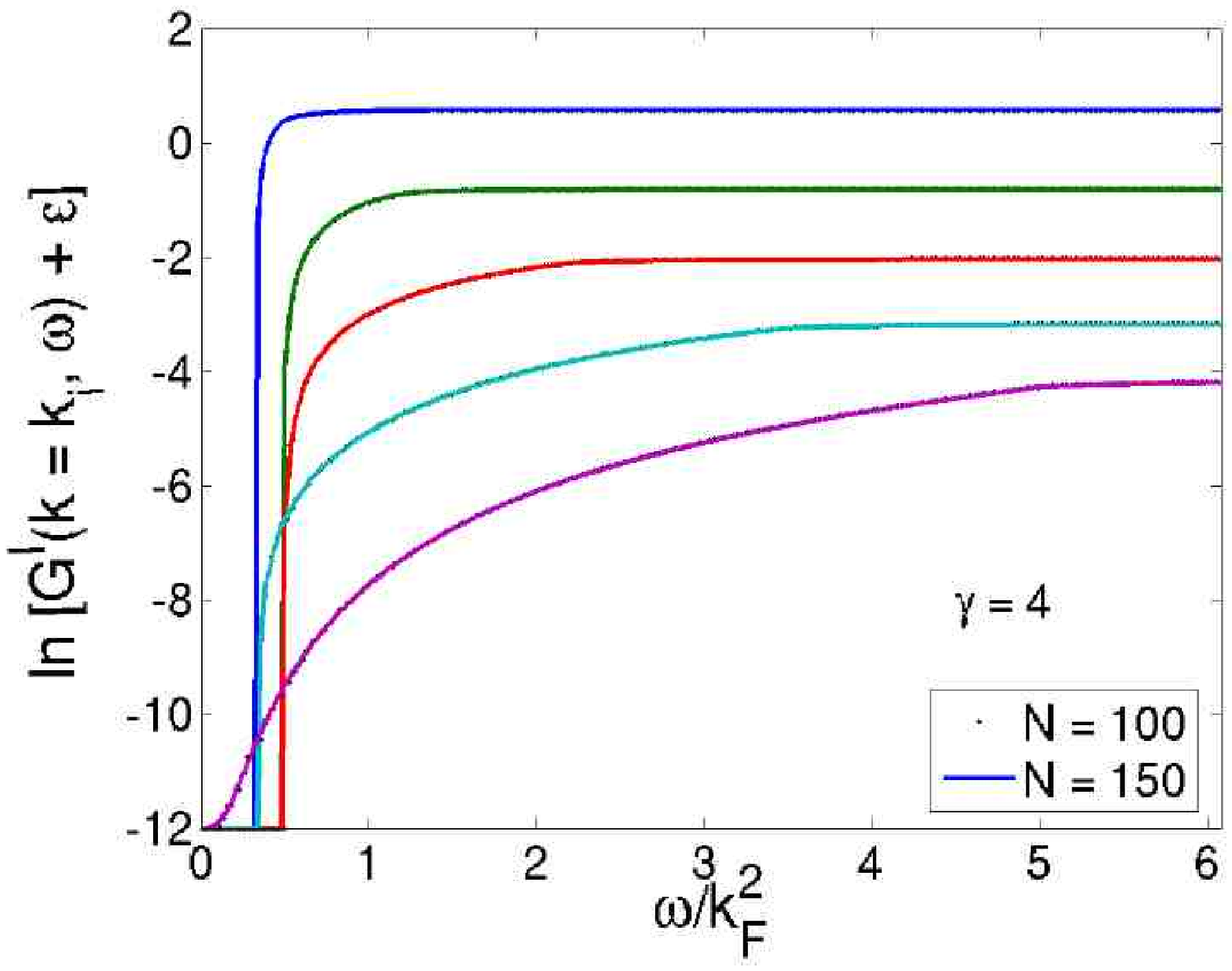} 
\end{tabular}
\label{GFk1}
\caption{Left:  fixed-momentum cuts of the logarithm of the one-body correlation function, 
for $\gamma = 1/4, 1$ and $4$.  As $\gamma$ increases, the peaks widen.  Right:  logarithm of
the integrated correlation function (\ref{GI}) for the same values of $k$ and $\gamma$  
($\varepsilon = 10^{-12}$ is a regulator for the logarithm).}
\end{figure}
\begin{figure}
\begin{tabular}{ll}
\includegraphics[width=7cm]{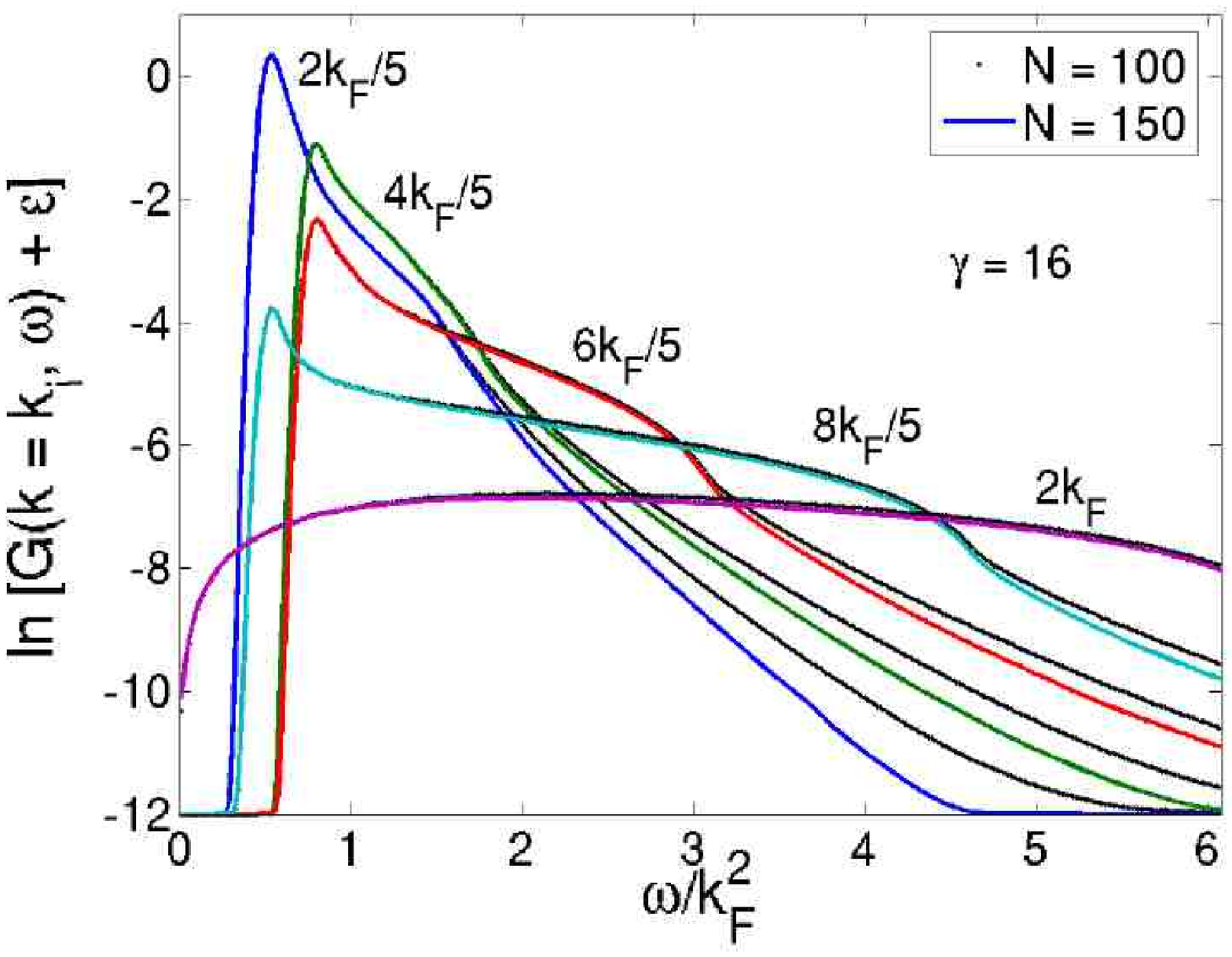} &
\includegraphics[width=7cm]{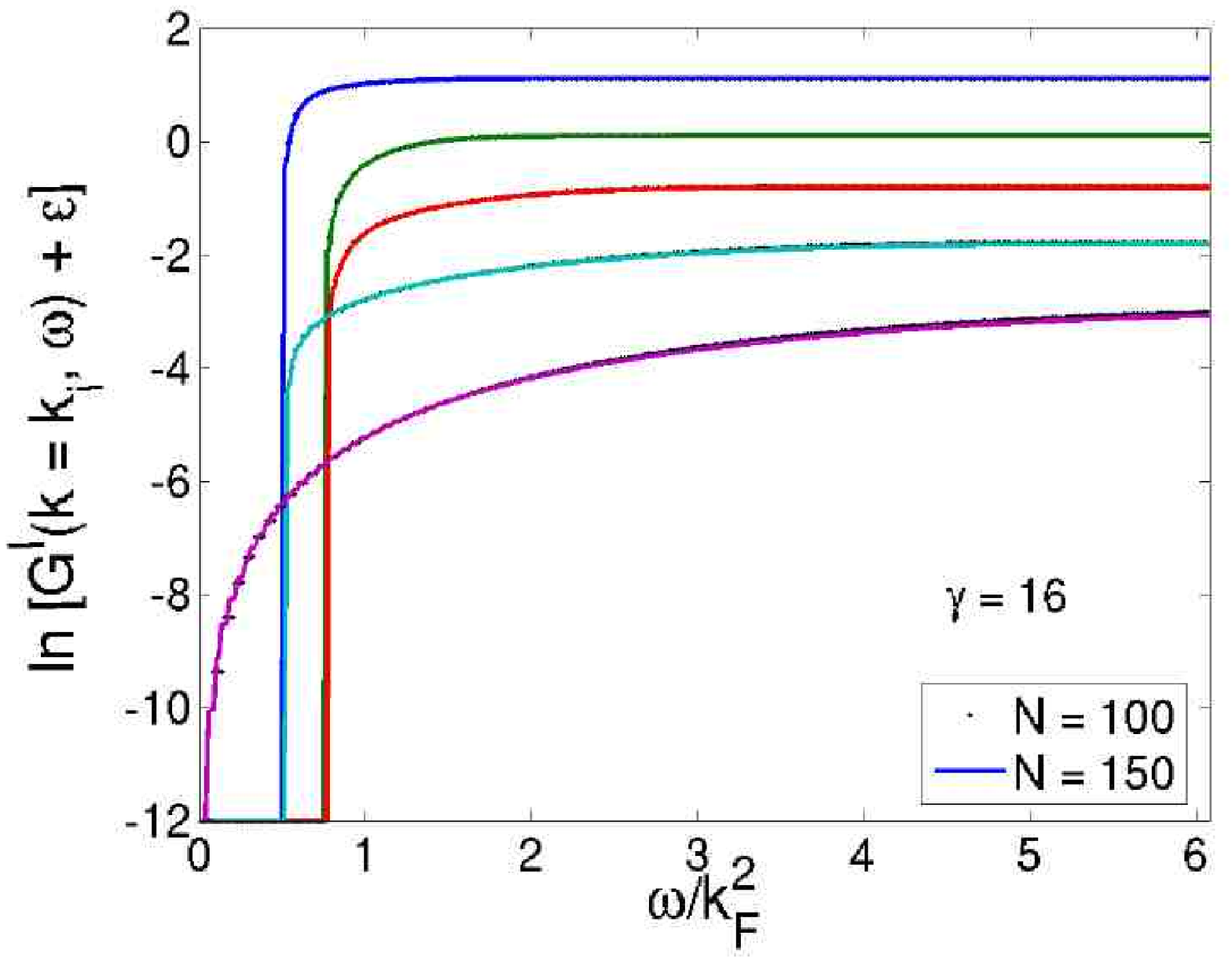} \\
\includegraphics[width=7cm]{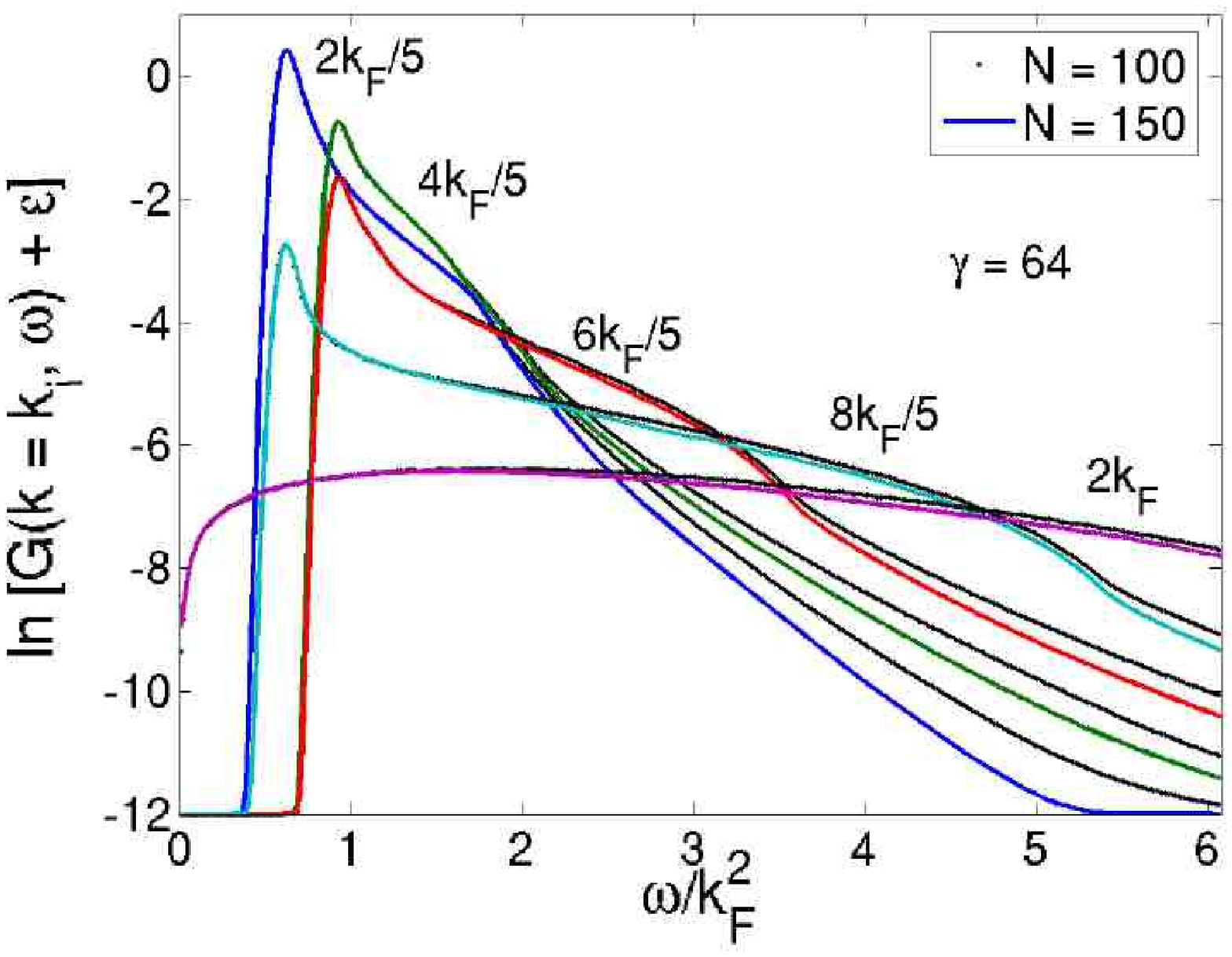} &
\includegraphics[width=7cm]{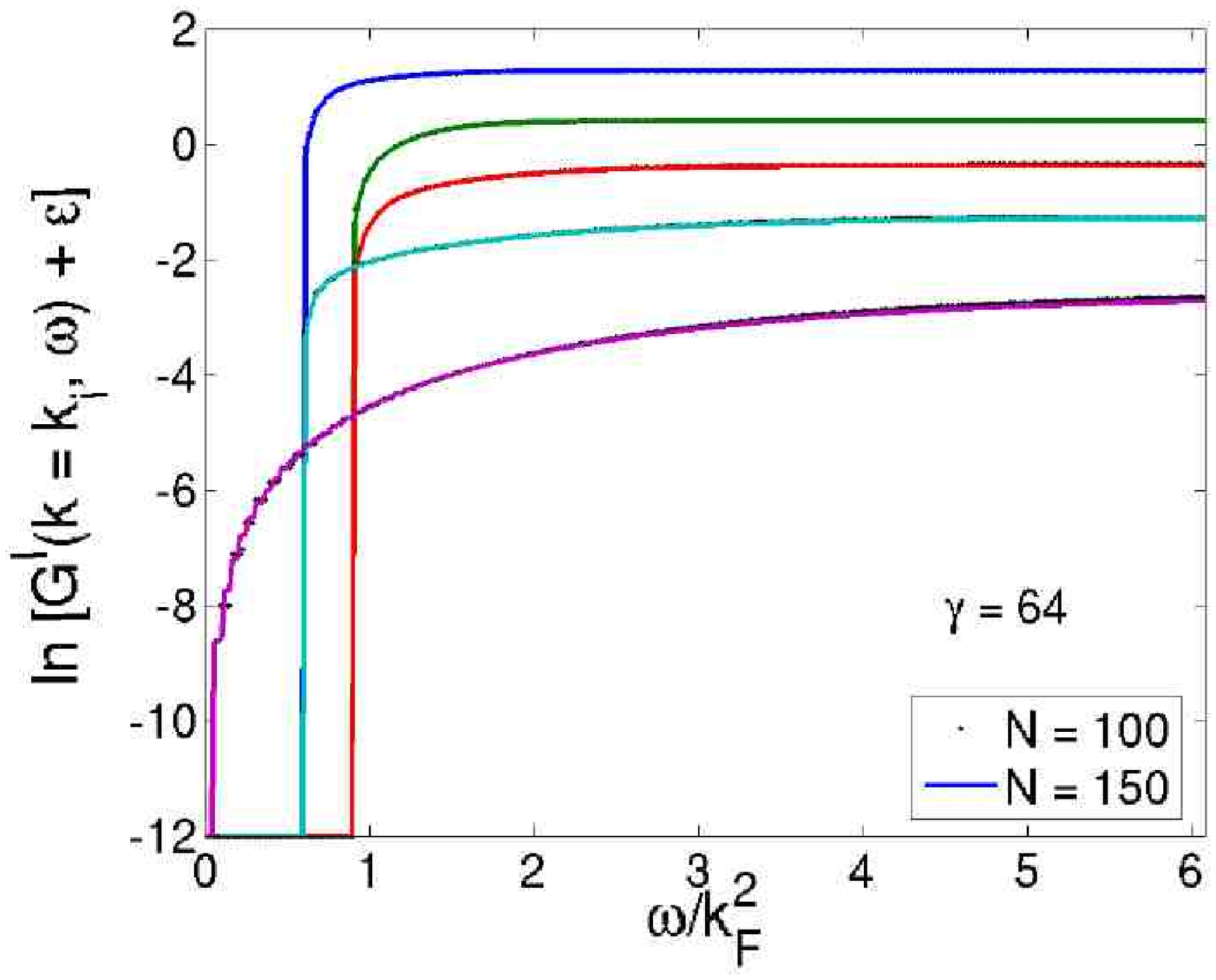} 
\end{tabular}
\label{GFk2}
\caption{Same as above, now for $\gamma = 16$ and $64$.  The peaks have moved to low energy.}
\end{figure}

\section{Momentum distribution function and one-body density matrix}

From the data of the dynamical one-particle correlation function
presented in the previous section it is very easy to recover the
static correlation function
\begin{equation}
n(k)=\int_0^{\infty} \frac{d\omega}{2\pi} G_2(k,\omega)\,.
\end{equation}
$n(k)$ represents the relative occupation of the single particle
state of momentum $k$ and is consequently known as the momentum
distribution function. It is the quantity that is directly measured
in the ballistic expansion of a trapped gas (see e.g., the review \cite{k-99}).
Its Fourier transform is the one-body density matrix
\begin{eqnarray}
\rho(x)&=& \sum_{s=-\infty}^\infty n\left(\frac{2\pi s}{L}\right)
\cos\left(\frac{2\pi x s}{L} \right)\\&=&
\frac{\int_0^L
\Psi^*_0(x_1+x,x_2,\dots,x_N)\Psi_0(x_1,x_2,\dots,x_N) dx_2 \dots
dx_N}{\int_0^L |\Psi_0(x_1,x_2,\dots,x_N)|^2 dx_1 \dots dx_N}\,,
\label{rhodef}
\end{eqnarray}
where $\Psi_0(x_1,x_2,\dots,x_N)$ is the $N$-body ground-state
wavefunction. $\rho(x)$ is the reduced density matrix of a single
particle when all the other degrees of freedom have been integrated
out.

Given its theoretical and experimental importance, the momentum
distribution function is the correlation function that has been
mostly studied in the literature. In the TG limit, its exact
expression is known since the work of Lenard
\cite{LenardJMP5}: for finite $N$ and $L$, $\rho(x)$
can be written as a simple Toeplitz determinant (see also
\cite{ffgw-03}) and the momentum distribution function can be
readily obtained via Fourier transform. The Lenard formula has
been re-obtained in the framework of algebraic Bethe Ansatz in Ref.
\cite{KorepinCMP129}. Some analytical asymptotic expansions in the TG limit
have been reported in Refs. \cite{asym,JimboPD1,ffgw-03,mvt-02}
and a $1/\gamma$ expansion has been developed in \cite{1gam}. We mention
that these results have been recently generalized to a lattice of
impenetrable bosons \cite{gs-06} and to anyonic gases \cite{ssc-06}.

\begin{figure}
\begin{tabular}{cc}
\includegraphics[width=7cm]{k4v3.eps}
&\includegraphics[width=7cm]{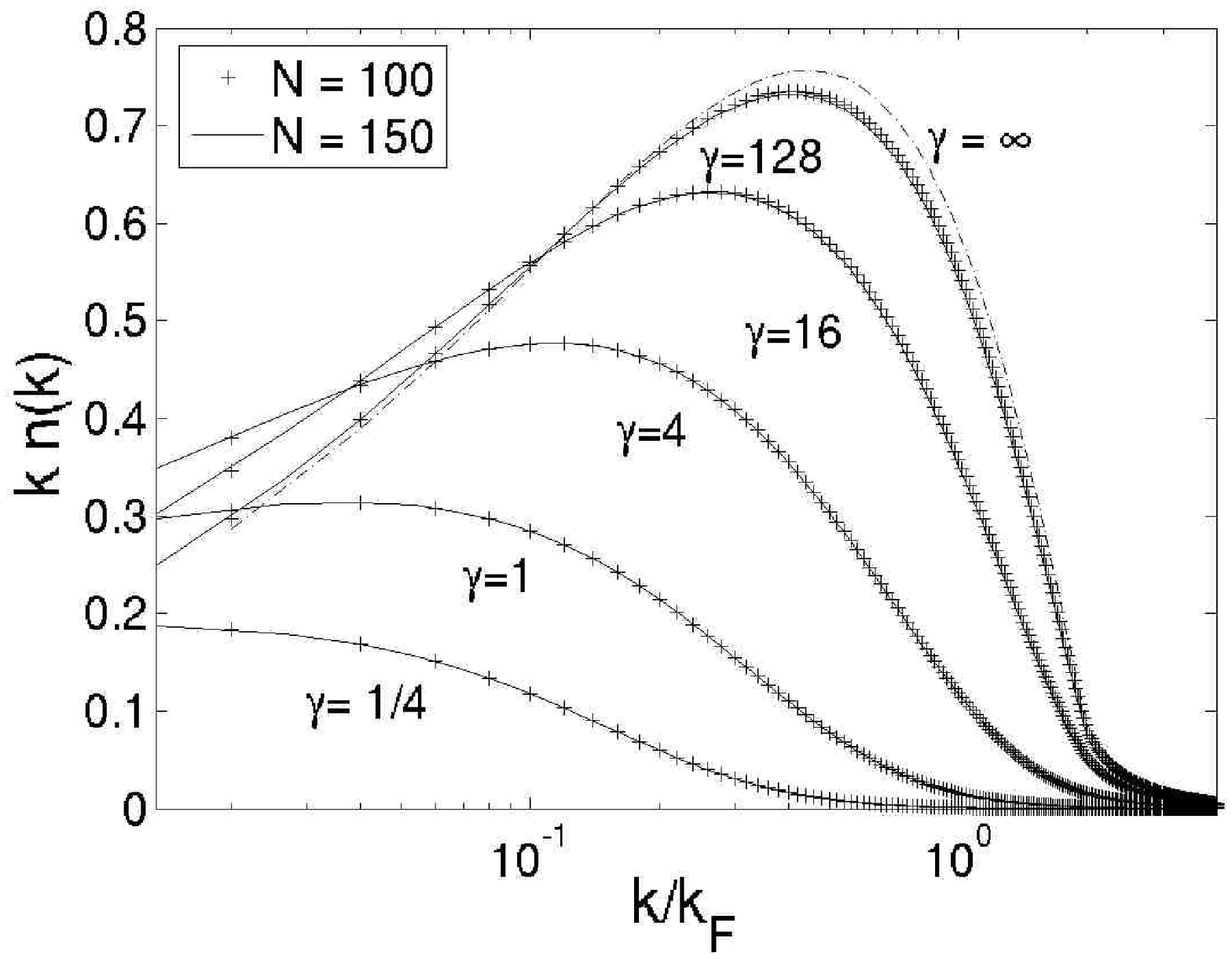}
\end{tabular}
\caption{
Momentum distribution function for some representative value of $\gamma$.
Left: $n(k)$ in log-log scale. In the inset the
large momentum section of and the asymptotic $k^{-4}$ law.
Right:  plot of $k n(k)$ as function of $k$ in log-scale
(this can be conveniently compared with the data of Ref. \cite{ag-03}).  
The exactly known
Tonks-Girardeau ($\gamma \rightarrow \infty$) limit is plotted as reference.
}
\label{SGF}
\end{figure}

In the case of finite coupling constant, only the asymptotic
expansion for small and large momenta are known.
The infrared behavior can be obtained from the hydrodynamic theory
of low-energy excitations \cite{ht}
\begin{equation}
n(k)\propto  k^{\alpha -1}\,,
\end{equation}
where $\alpha=v_s/(4\pi n)$ and $v_s$ is the speed of sound. The above
result holds for $k\ll \xi^{-1}$, where $\xi=\sqrt2/v_s$ is the
healing length. The corresponding behavior of the one-body density
matrix is $\rho(x)\propto x^{-\alpha}$ for $x\gg\xi$. The exponent
$\alpha$ interpolates between $0$ and $1/2$ when the coupling
constant $\gamma$ is increased from $0$ to $\infty$.
The large momentum tail can be elegantly obtained by using simple
theorems on the Fourier transform \cite{od-03} and can be related to
the previously calculated second moment of the density operator
\cite{GangardtPRL90}, yielding ($n(k)$ corresponds to $w(p)$ in 
Ref. \cite{od-03} and not to $W(p)$)
\begin{equation}
N n(k)= {\gamma^2 e'(\gamma)}\left(\frac{n}k\right)^4\,,
\qquad {\rm for}\,\; k\to\infty
\label{k-4}
\end{equation}
where $e(\gamma)$ is the ground-state energy per particle in the thermodynamic
limit (in the TG limit this formula was previously obtained in Ref.
\cite{mvt-02,ffgw-03}). 
Similarly a small $x$ expansion can be written for the one-body density
matrix:
\begin{equation}
\frac{\rho(x)}n=1+\sum_{i=1}^\infty c_i |n x|^i\,.
\label{smallxeq}
\end{equation} 
The $k^{-4}$ law for $n(k)$ implies that the lowest odd power in this Taylor
expansion is $|x|^3$, thus $c_1=0$. The coefficient $c_3$ is easily read 
from Eq. (\ref{k-4}), yielding $c_3=\gamma^2 e'(\gamma)/12$ \cite{od-03}. 
$c_2$ can be obtained \cite{od-03} using the 
Hellmann-Feynman theorem: $c_2=-1/2[e(\gamma)-\gamma e'(\gamma)]$.
The other coefficients $c_i$ with $i>3$ are still unknown.


To the best of our knowledge the results reported above are all that is
exactly known about the momentum distribution function (some results for 
finite and small $N$ are reported in Ref. \cite{ffm-06}). In
particular nothing is known for finite coupling in the intermediate
window of momenta, that is actually the region easier to access in
experiments (in fact the infrared behavior is usually obscured by
the harmonic trap, see 
e.g. \cite{ParedesNATURE429,sf-06,p-03,ag-03,LuxatPRA67}). 
As a consequence, up to now this interesting regime has been investigated 
only with purely numerical methods as Quantum Monte Carlo \cite{ag-03}
and density matrix renormalization group \cite{sf-06}.
However, these methods, although very powerful, were unable to describe
all the physics from the infrared to the ultraviolet.
Our results demonstrate that the ABACUS method is able to do this.

In Fig. \ref{SGF} we present $n(k)$ as function of $k$ for
several values of the interacting parameter and for $N=100,150$.
The curves clearly show that finite size effects are negligible on
this scale. 
We also plot $k n(k)$ for comparison with the Quantum Monte Carlo 
data of Ref. \cite{ag-03}: the data shows an overall qualitative agreement.
The exact result of Lenard for the TG limit is also given for comparison.

One advantage of our method is that
we are able to reproduce the $k^{-4}$ law to high accuracy. 
In the inset of Fig. \ref{SGF} we plot $n(k)$ versus $k$ for large momenta 
and for several
values of $\gamma$.  The $k^{-4}$ tails are evident, and agree with the
asymptotic curves predicted from Eq. (\ref{k-4}). We
stress that in these curves there is no fitting parameter, both the
power-law and the numerical prefactor are fixed according to Eq. (\ref{k-4}).
This therefore confirms that our method provides accurate results
even at high momentum.

\begin{SCfigure}
\includegraphics[width=8cm]{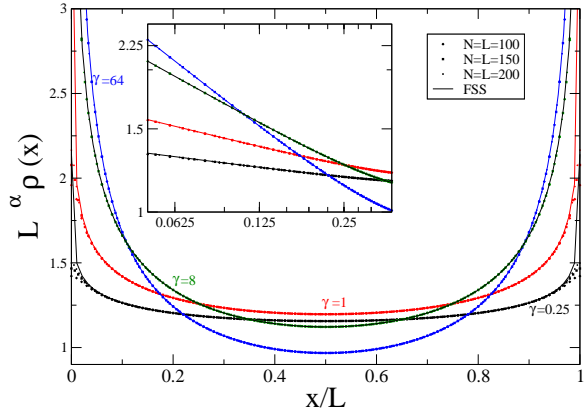}
\vspace{3mm}
\caption{One-body density matrix $\rho(x)$ 
for $\gamma=64,8,1,0.25$ and $N=100,150,200$ compared with the finite-size 
scaling (FSS) form.  
Inset: The same plot in log-log scale to show the power law 
behavior for $\xi\ll x< L/2$.}
\label{rhoFSS}
\end{SCfigure}

Now we turn to the real spatial dependence of this correlation function.
From our data, the one-body density matrix is easily obtained by
means of Fourier transform according to Eq. (\ref{rhodef}). 
In Fig. \ref{rhoFSS} we plot $\rho(x)$ as function of $x/L$ for some
values of $\gamma$. The inset in log-log scale clearly shows the power-law 
behavior $x^{-\alpha}$ for $x\gg\xi$ (but smaller than $L/2$).
At large distances there are evident deviations due to the 
finiteness of the system.
Conformal field theory \cite{cardyb} predicts that for
periodic boundary conditions the correlation function gets modified 
according to
\begin{equation}
\rho(x)\propto\left(\frac{\pi}{L \sin (\pi x/L)}\right)^\alpha\,.
\label{rhoFSSeq}
\end{equation}
For $x\ll L$, this form reproduces the power-law $x^{-\alpha}$, but it 
also completely describes the behavior for any $x$ with the only condition
$x\gg\xi$. In fact, plotting $L^\alpha \rho(x)$ versus $x/L$ 
all the curves for various values of $N$ fall on the same master curve, 
which depends on $\gamma$ only through the exponent $\alpha$. 
The proportionality constant in Eq. (\ref{rhoFSSeq}) can be fixed
by imposing a given value at $x=L/2$, but in Fig. \ref{rhoFSS} we let it free
and consequently there is no fitting parameter again.
It is evident that only for very small value of $x/L$ it is possible to 
notice the effect of the finite healing length $\xi$.

\begin{figure}[b]
\includegraphics[width=11cm]{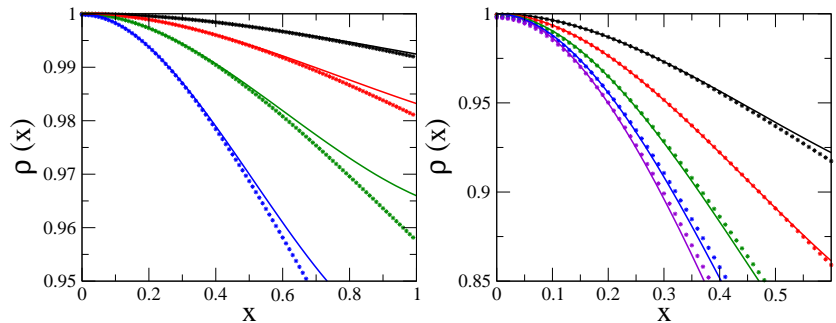}
\caption{Small $x$ behavior of the one-body density matrix $\rho(x)$,
compared with the asymptotic result Eq. (\ref{smallxeq}).
Left: $\gamma=0.25,0.5,1,2$ from top to bottom.
Right:   $\gamma=4,8,16,32,64$ from top to bottom.
}
\label{smallxrho}
\end{figure}

The small $x$ behavior is also easily accessed with ABACUS. 
In Fig. \ref{smallxrho} we report $\rho(x)$ for very small $x$ (up to 
$x=1$) for $N=100$. Compared to Fig. \ref{rhoFSS} we are now showing
a very small fraction of the total plot. The numerical data are compared 
with the Taylor expansion Eq. (\ref{smallxeq}) with the coefficients $c_2$
and $c_3$ evaluated in Ref. \cite{od-03}. 
For very small $x$ the agreement is perfect and again there is no fitting 
parameter.
However, we notice that Eq. (\ref{smallxeq}) with only two $c_i$'s
describes a very limited region, probably very difficult to reach in 
real systems.
A curious fact can be easily deduced from Fig. \ref{rhoFSS}:  the 
correction to the asymptotic result is positive for large $\gamma$ and negative
for small values, so the unknown coefficient $c_4$ changes sign as $\gamma$ increases
(contrary to $c_2$ and $c_3$, which are always respectively negative and
positive).  From our data, this change should occur around $\gamma\sim 8$.

\section{Conclusion}

In conclusion, we have presented results for the one-body dynamical correlation function of the 
one-dimensional Bose gas (Lieb-Liniger model) by using the ABACUS method, mixing integrability and 
numerics.  We obtained the full momentum and frequency dependence of the correlations, for values
of the interaction parameter interpolating continuously from the weakly-interacting regime to the
strongly-interacting Tonks-Girardeau regime.  We wish to stress that the present method not
only yields previously inaccessible results on the dynamics, but by extension also allows to go
beyond other methods when restricting to static quantities.  For example, we have showed how our results, 
integrated to recover the static correlation function, are the first to be in agreement with all 
asymptotic predictions.  

We hope that our results will provide motivation for further
experimental work on quasi-one-dimensional gases, in particular on
the direct observation of dynamical correlation functions in these systems.
On the other hand, we intend to use the correlation functions we have obtained for 
the idealized Bose gas as a starting point for addressing more general situations,
including the effects of intertube coupling and/or confining potential.  Other
possibilities include the study of entanglement in this stongly-correlated system, using measures expressible in terms of 
simple combinations of the correlation functions we have studied.  We will report on these issues in future
publications.

\section*{Acknowledgments}
J.-S. Caux and P. Calabrese acknowledge support from the Stichting voor Fundamenteel Onderzoek der Materie (FOM) in the Netherlands.
The computations were performed on the LISA cluster of the SARA computing facilities in the Netherlands.
N. S. is supported by the French-Russian Exchange Program, the Program of RAS Mathematical Methods of Nonlinear Dynamics,
RFBR-05-01-00498, SS-672.2006.1.
J.-S. C. and P. C. would like to acknowledge interesting discussions with D. M. Gangardt.

\end{document}